\documentclass [12pt]{article}
\pdfoutput=1
\usepackage{graphicx}
\usepackage{hyperref}
\hypersetup{colorlinks,citecolor=black,linkcolor=black,urlcolor=black}

\begin{document}

\title{Graphene and Boron Nitride Single Layers}

\maketitle

\noindent
\noindent
\linebreak[4]
\textnormal{Thomas Greber, Physik Institut, Universit\"at Z\"urich, Switzerland\\
Uni Irchel 36K88,
+41 44 635 5744, greber@physik.uzh.ch}
\\
\\
\linebreak[4]
\texttt{Chapter in the Handbook of Nanophysics, Taylor  and Francis Books, Inc.  {Editor Klaus Sattler}}
\vspace{0.5cm}

\newpage

\tableofcontents
%\bigskip

\newpage

This Chapter deals with single  layers of carbon (graphene) and hexagonal boron nitride on transition metal surfaces. 
The transition metal substrates take the role of the support and allow due to their catalytic activity the growth of perfect layers by means of chemical vapor deposition. 
The layers are $sp^2$ hybridized honeycomb networks with strong in plane $\sigma$ and weaker $\pi$ bonds to the substrate and to the adsorbates.
This hierarchy in bond strength causes anisotropic elastic properties, where the $sp^2$ layers  are stiff in plane and soft out of plane. 
A corrugation of these layers imposes a third hierarchy level in bond energies,  with lateral bonding to molecular objects with sizes between 1 and 5 nanometer.
This extra bond energies are in the range of thermal energies $k_BT$ at room temperature and are particularly interesting for nanotechnology.
The concomitant template function will be discussed in Section \ref{functextured}.
The peculiar bond hierarchy also imposes intercalation as another property of $sp^2$ layer systems (Section \ref{intercalation}).
Last but not least $sp^2$ layer systems are particularly robust, i.e. survive immersion into liquids \cite{wid07}, which is a promise for $sp^2$ layers being useful outside ultra high vacuum.

The Chapter shortly recalls the synthesis, describes the atomic and electronic structure, is followed by a discussion of properties like intercalation and the use of $sp^2$ layers on metals as tunneling junctions or as templates.
The Sections are divided into subsections along the sketch in Figure \ref{F1}, i.e. for flat and corrugated layers. Since there are flat and corrugated layers for graphene (g) as well as hexagonal boron nitride (h-BN) the similarities and differences between C-C and B-N are discussed in every subsection. 
The Chapter ends with an Appendix that summarizes the basics of atomic and electronic structure of honeycomb lattices.

Of course the Chapter does not cover all aspects of $sp^2$ single layers.
Topics like free standing layers \cite{gei07}, edge structures of ribbons \cite{cer08, eno07}, topological defects \cite{cor08}, or mechanical and chemical properties were not  covered.
\newpage
\section{Single layer systems}

\begin{figure}
	\begin{center}
	\includegraphics[width=0.7\textwidth]{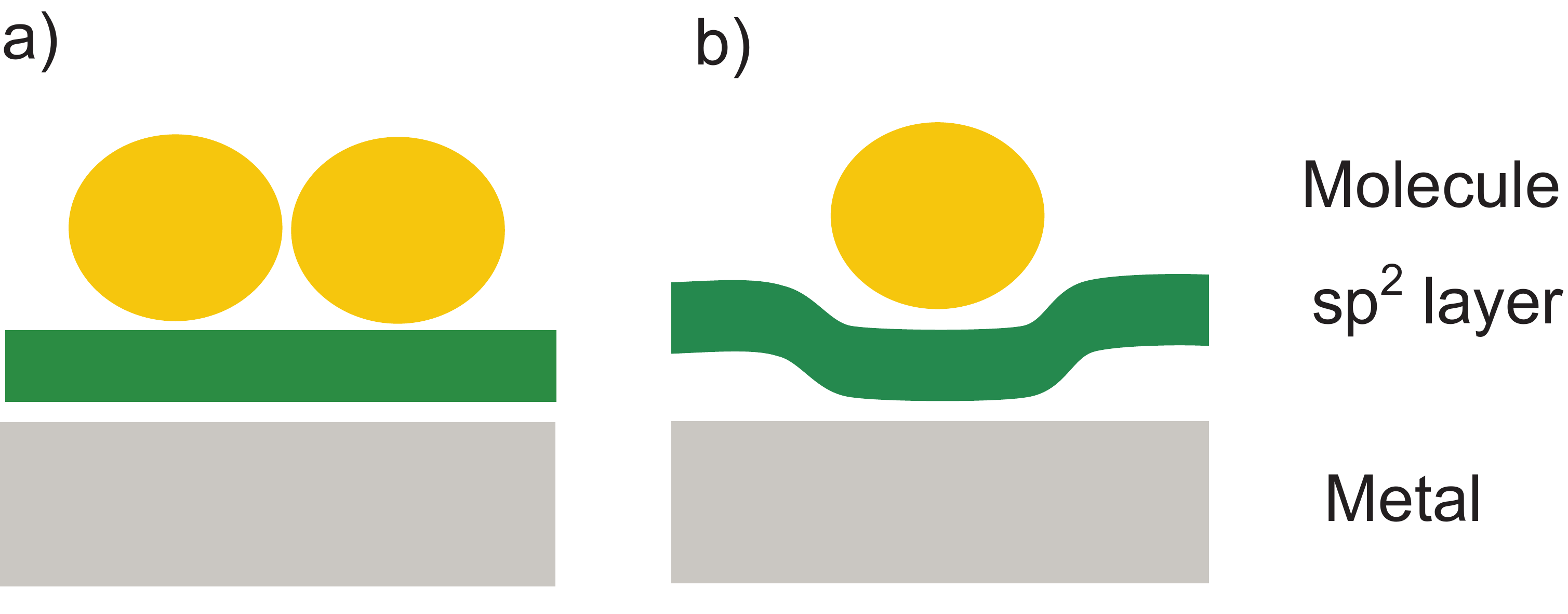}
		\caption{Schematic side view of a single layer on top of a metal. The layer changes the properties of adsorbed molecules as compared to the pristine surface.
a) Flat layer, where molecule-molecule interactions are present (see Figure \ref{FSTMNi} and \ref{c60hbnni}.  b) Corrugated layer, which is a template on which single molecules can be laterally isolated (see Figure \ref{FSTM}).}
		\label{F1}
	\end{center}
\end{figure}

A single layer of an adsorbate strongly influences the physical and chemical properties of a surface.
Sticking and bonding of atoms and molecules may change by orders of magnitude, as well as the charge transport properties across and parallel to the interface.

There are many  single layer systems like e.g.  graphite/graphene \cite{she74,ndi06}, hexagonal boron nitride \cite{paf90,auw99},  boron carbides \cite{yan04}, molybdenum disulfide \cite{hel00}, sodium chlorides \cite{ben99,piv05}, or aluminium oxide \cite{nil08}, copper nitride \cite{bur76,rug07} to name a few. 
In order to decide whether single layers are "dielectric" or "metallic" the electronic structure at the Fermi level has to be studied, where a metallic layer introduces new bands at the Fermi energy, while a dielectric layer does not.

Figure \ref{F1} shows a schematic view of the single layer systems, with adsorbed molecules. The beneath conduction electrons of the substrate still have tunneling contact to the molecule, though the bonding is much weaker compared to the bare metal. If the single layer is flat (Figure \ref{F1} a)) adsorbed molecules may touch, while in corrugated  layer systems (Figure \ref{F1} b)) molecules may be trapped also laterally on the surface. 
The corrugation (see Secton \ref{atomictextured}) results from the epitaxial stress due to the lattice misfit between the substrate and the $sp^2$ layer.
This results in lateral superstructures with nanometer periodicities and corrugations in the 0.1 nm range.
For developments in nanotechnology it is useful to have single layer systems which are inert, remain clean at ambient conditions and are stable up to high temperatures. 
In this field sp$^2$ hybridized graphene and hexagonal boron nitride are outstanding examples.
These two intimately related model systems, are discussed in more detail.
Both are $sp^2$ hybridized layers with about the same lattice constant, but, while on most transition metals graphene is metallic, h-BN is an insulator.
\newpage
\section{$sp^2$ single layers}

In this Chapter we restrict ourselves to the $sp^2$ hybridized single layer systems of graphene and hexagonal boron nitride.
Except for some considerations in the Appendix we always deal with layers on transition metal supports.
The comparison between graphene and boron nitride is particularly helpful for the understanding of their functionality.
They are siblings with similarities and differences.
Both form a $sp^2$ honeycomb network with similar lattice constant (0.25 nm), with strong $\sigma$ in plane bonds and soft $\pi$ bonds to the substrate and the adsorbates.
Depending on the substrate, both form flat or corrugated overlayers.
The corrugation is a consequence of the lattice mismatch and the anisotropic bonding of the atoms in the honeycomb with the substrate, and leads to the peculiar functionality of molecular trapping \cite{dil08}.
Graphene and h-BN differ in their atomic structure: In the case of h-BN two different atoms constitute the honeycomb unit cell, while the two carbon atoms are equivalent within the honeycomb lattice.
This causes most graphene overlayers to be metallic, while the h-BN layers are insulators.
It is furthermore the reason for an "inverted topography" if corrugation occurs (see Figure \ref{F1Brugger}).

The theoretical description of $sp^2$ single layers \cite{wal47} is easier than the experimental realisation \cite{nov05}.
The first experimental reports on $sp^2$ layer production on supports dates back to early times of surface science \cite{kar66} and has sometimes been considered to be an annoyance since they poison catalysts \cite{she74}.
In the field of ultra-thin epitaxial films of graphene and hexagonal boron nitride on solid surfaces the work of the Oshima group has to be highlighted \cite{osh97}. 
They studied the production of $sp^2$ layer systems systematically and carefully characterized their electronic and vibronic structures. 
Today this work is the starting point for numerous ongoing investigations, also because of the rise of graphene \cite{gei07} i.e. the discovery that devices containing free-standing single layer devices may be realized. Therefore there is a strong demand for production routes for these materials that have, somehow, to start on a solid surface. 
 
\subsection{Synthesis}
\label{synthesis}

There are several roads to the production of $sp^2$ single layers. 
Here the Chemical Vapor Deposition (CVD) processes that base on the reactive adsorption of precursor molecules from the gas phase onto the substrate, and the segregation method, where the constituents diffuse from the bulk to the surface,  are briefly summarized.

The $sp^2$ layers may also be grown on substrates without C$_{3v}$ symmetry as e.g. present for the (111) surfaces of face centered cubic $fcc$ crystals.
Examples are the growth of h-BN on Mo(110) \cite{all07} or on Pd(110) \cite{cor05}.

Although this chapter deals with single $sp^2$ layers only, it could be desirable to grow multilayers. 
There is e.g. one report on the growth of graphene on h-BN/Ni(111) \cite{nag96}, and for the segregation approach it seems to be easier to grow multilayers, as it was shown for silicon carbide (SiC) substrates \cite{for98}, since this method has not to overcome the low sticking probability of precursor gases on complete $sp^2$ layers.

\subsubsection{Chemical Vapor Deposition (CVD)}

Chemical Vapor Deposition processes comprise the adsorption and reaction of the educts i.e. precursor molecules on the surface where a new material shall grow. 
The process often involves the cracking or decomposition of the precursor  molecules and a partial release of products into the gas phase.

\begin{figure}
\centerline {\includegraphics[width=0.5\textwidth]{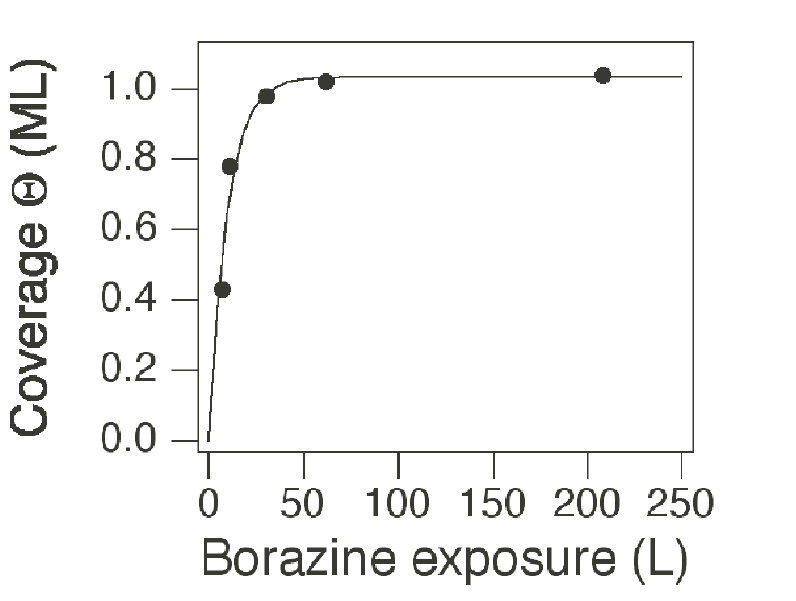}}
\caption{Growth of a h-BN layer on Ni(111): Coverage as a function of borazine exposure. 1 Langmuir (L)= $10^{-6}  Torr \cdot s$. Note the drop of growth rate after the completion of the first layer. By courtesy of W, Auw\"arter. }
\label{layergrowth}
\end{figure}

The first h-BN single layers have been synthesized on Ru(0001) and Pt(111) surfaces by CVD of benzene like borazine $(HBNH)_3$  \cite{paf90}. 
The process comprises the hydrogen abstraction from the borazine molecules, the assembly of hexagonal boron nitride and the desorption of $H_2$ gas.
The h-BN growth rate drops after the formation of the first layer by several orders of magnitude.
This has the practical benefit that it is easy to prepare single layers.
In Figure \ref{layergrowth} the growth of h-BN on Ni(111) is shown as a function of the exposure to the precursor molecules.  Clearly, the growth rate drops by more than two orders of magnitude after the completion of the first layer.
This behavior is quite general for $sp^2$ layer systems and similar growth behavior is expected for graphene layers.
The drop in growth rate is presumably due to the much lower sticking probability of borazine after the completion of the first h-BN layer, and the much lower catalytic activity of h-BN compared to clean transition metals. 
However, not much is known on the details of the growth process of h-BN.
From the study of h-BN island morphologies on Ni(111) it was suggested that the borazine BN six-ring is opened during the self assembly process \cite{auw03,auw04}. 
For h-BN layer formation, also trichloroborazine (ClBNH)$_3$ \cite{auw04}, and diborane (B$_2$H$_6$) ammonia (NH$_3$) gas mixtures \cite{des97} were successfully used.

For the formation of graphene layers many  different precursors have been used, as e.g. ethene ($C_2H_4$) \cite{ndi08}, acetylene ($C_2H_2$) \cite{nag96}, or propylene ($C_3H_6$) \cite{shi00}.
It turns out that almost all hydrocarbons react in non oxidizing environments on transition metal surfaces to graphene, though there are also growth conditions where diamond films grow \cite{gse08}.

\subsubsection{Segregation}

An alternative way to the above mentioned CVD processes is the use of segregation. Here the educts are diluted in the bulk of the substrate. If the substrate is heated, they start to diffuse and eventually meet the surface, where they have a larger binding energy than in the bulk. At the surface they may then react to the new material. 
If the substrate temperature exceeds the stability limit of the $sp^2$ layer, back dissolution into the bulk or desorption may occur.
Well known examples are e.g. the formation of graphene on Ni(111) \cite{she74,eiz79}, Ru(0001) \cite{mar07}, SiC \cite{for98} or the formation of BC$_3$ layers on NbB$_2$  \cite{yan04}.

%%%%%%%%%%%%%%
%%%%%%%%%%%%%%
\subsection{Atomic structure}

The atomic structure of the $sp^2$ layer systems is fairly well understood. 
It bases on a strong $sp^2$ hybridized in plane bonded honeycomb lattice and a, relative to these $\sigma$ bonds, weak $\pi$ bonding to the substrate.
The $\pi$ ($p_z$) bonding depends on the registry to the substrate atoms, where the layer-substrate ($p_z-d_{z^2}$) hybridization causes a tendency for lateral lock-in of the overlayer atoms to the substrate atoms.
For systems, where substrates have the same symmetry ($C_{3v}$), as the $sp^2$ honeycombs, the lattice mismatch ${M}$ is defined as:

\begin {equation}
{M}=\frac{a_{ovl}-a_{sub}}{a_{sub}}
\label{mismatch}
\end{equation}

where $a_{ovl}$ is the lattice constant of the overlayer, and $a_{sub}$ that of the substrate. 
Often the sign of the lattice mismatch is not indicated, and the absolute value of the difference between the substrate and the overlayer lattice constant is used. 
Then it has to be explicitly said whether the mismatch induces compressive (+) or tensile (-) stress in the overlayer, or vice versa, tensile (+) or compressive (-) stress in the substrate.
The mismatch plays a key role in the epitaxy of the $sp^2$ layers. 
Of course, the bonding to the substrate also depends on the substrate type, and in turn may lead to the formation of corrugated super structures with a peculiar template function.
\begin{figure}
	\centerline{\includegraphics[width=0.5\textwidth]{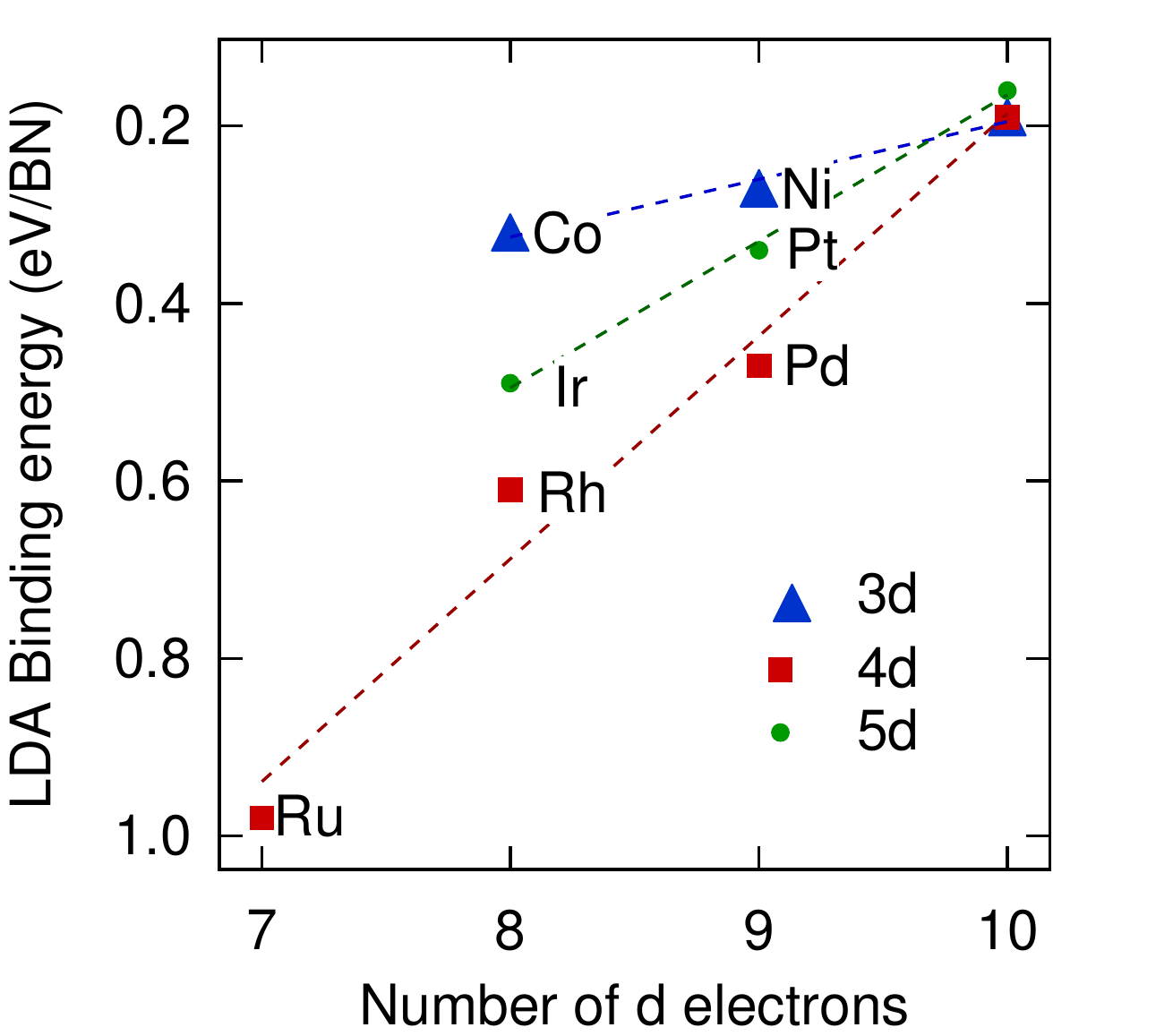}}
	\caption{Calculated BN binding energies of h-BN on various transition-metals. The h-BN is strained to the lattice constant of the substrate. The bond strength correlates with the $d$-band occupancy, where the bonding is strongest for the 4d metals. The dashed lines are guides to the eyes. Results within the Local Density Approximation (LDA) are shown. Data from \cite{las08}.}
\label{FLaskowski}
\end{figure}

Figure \ref{FLaskowski}  shows the calculated BN bond strength on different transition metals (TM) \cite{las08}. The calculations were performed for (1x1) unit cells, where the h-BN was strained to the lattice constant of the corresponding substrate, with nitrogen in on top position. 
The BN bond energy is determined from the difference of the energy of the (1x1) h-BN/TM system and the strained h-BN  + TM system.
The trend indicates the importance of the $d$-band occupancy of the substrate and it is e.g. similar to the dissociative adsorption energies of ammonia $NH_3$ on transition metals \cite{bli04}. 
Since the lock-in energy, i.e. the energy that has to be paid when the nitrogen atoms are moved laterally away from the on top sites, is expected to scale with the adsorption energy, Figure \ref{FLaskowski} also rationalizes e.g. why h-BN/Pd(111) is less corrugated than h-BN/Rh(111).

This section is divided into two parts, where we distinguish flat and corrugated layers. 
Flat means that the same types of atoms have the same height above the substrate, which has only to be expected for (1x1) unit cells like that in the h-BN/Ni(111) system. 
Corrugation occurs due to lattice mismatch and lock-in energy gain, i.e. due to the formation of dislocations.

%%%%%%%%%%%%%%%%%%%%%%%%
\subsubsection{Flat layers: Domain boundaries}
\label{atomicflat}
\begin{figure}
	\centerline{\includegraphics[width=0.8\textwidth]{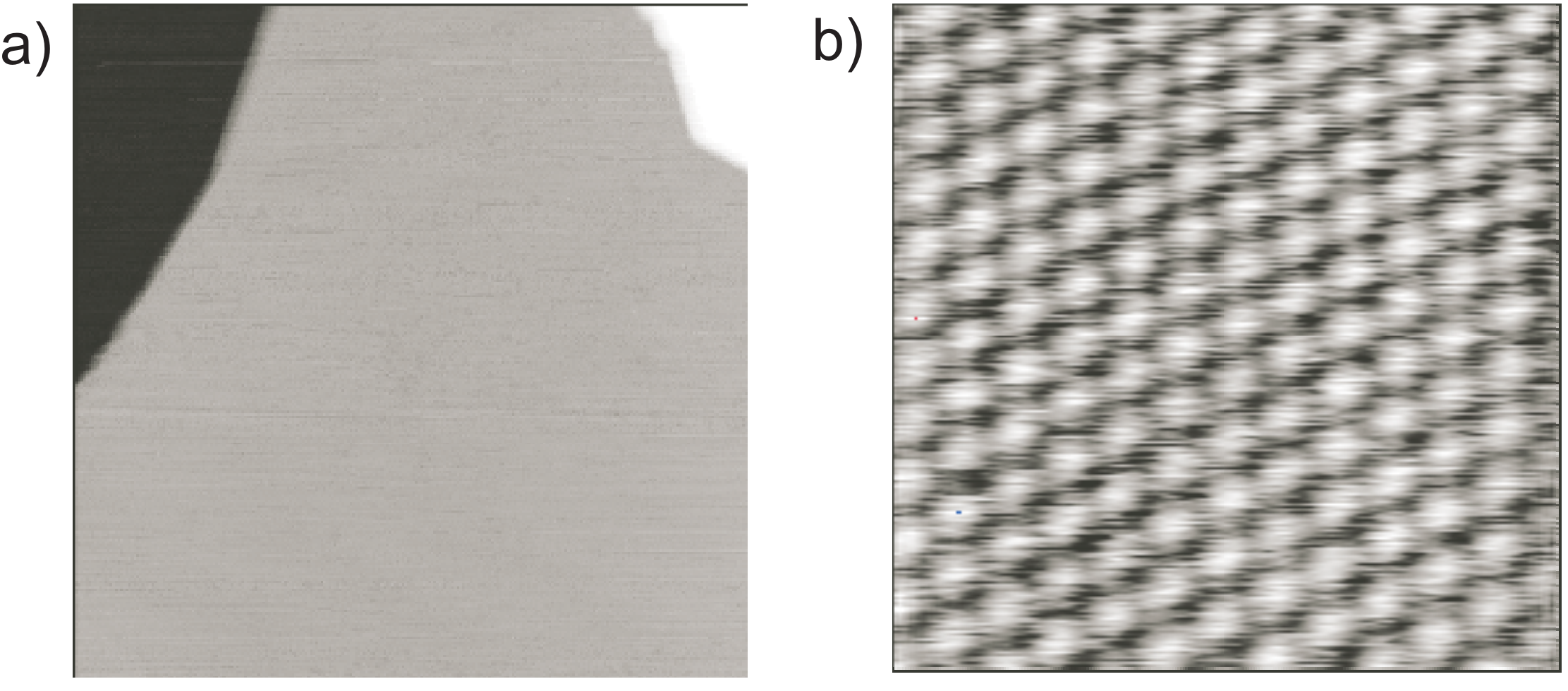}}	
		\caption{ Constant current scanning tunneling microscopy (STM) images of a flat layer of hexagonal boron nitride on Ni(111). a) 30x30 nm. The grey scales indicate 3 terraces with different heights. Large defect free terraces form. b) 3x3 nm. The nitrogen (bright)  and the boron (grey) sublattices are resolved with different topographical contrast. From \cite{auw99}.}
		\label{FSTMNi}
\end{figure}

The Ni(111) substrate has $C_{3v}$  symmetry  and a very small lattice mismatch of +0.4 \% to the $sp^2$ layers.
On Ni(111) h-BN forms perfect single layers.
Figure \ref{FSTMNi} shows scanning tunneling microscopy (STM) images from h-BN/Ni(111).
The layers appear to be defect free and flat. 
The STM may resolve the nitrogen and the boron sublattices, where a Tersoff Haman calculation indicated that the nitrogen atoms map brighter than the boron atoms \cite{gra03}. 
The production process of the h-BN layers (see Section \ref{synthesis}) also leads to the formation of larger terraces compared to uncovered Ni(111), where widths of 200 nm are easily obtained.
The stability of the (111) facet was also observed in experiments with stepped Ni(755) \cite{rok99} and Ni(223) surfaces, where the miscut of these surfaces relative to the [111] direction lead to large (111) facets and step bunches.

\begin{figure}
	\centerline{\includegraphics[width=0.8\textwidth]{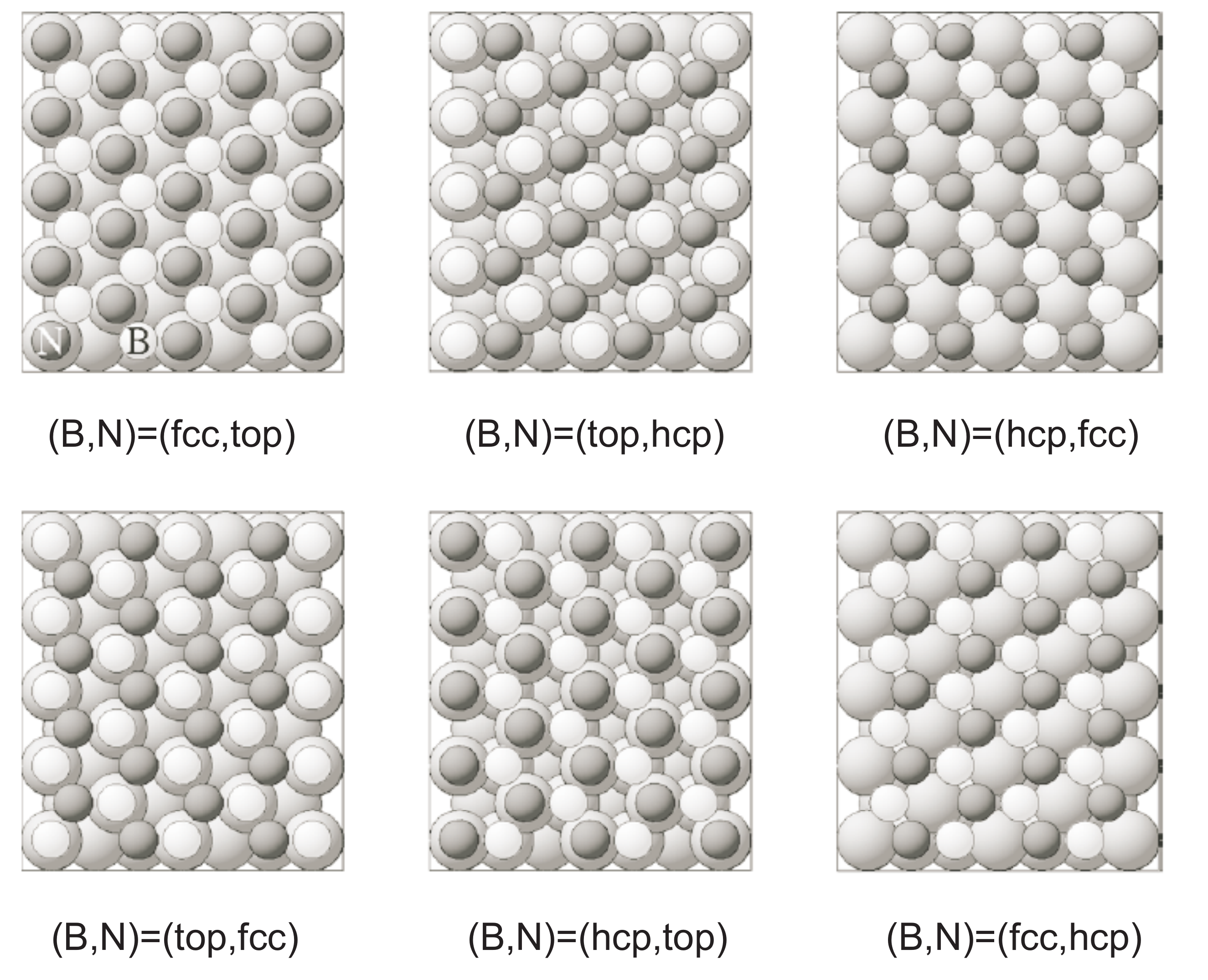}}
	\caption{Possible (1x1) configurations for one-monolayer h-BN/Ni(111). The top, fcc hollow, and hcp hollow sites are considered
for the position of the boron and nitrogen atoms, respectively. The (B,N)=(fcc,top) registry has the strongest bond, while calculations predict 9 meV lower binding energy for the  (B,N)=(hcp,top) configuration. From \cite{gra03}.}
\label{FGrad}
\end{figure}

For the case of the h-BN/Ni(111) system the mismatch ${M}$ is +0.4\%, i.e. the h-BN is laterally weakly compressed. This small mismatch leads to (1x1) unit cells and atomically flat layers.
The atomic structure of h-BN/Ni(111) is well understood and there is good agreement between experiment \cite{gam97,auw99,mun01} and theory \cite{gra03,che05,hud06}.
Figure \ref{FGrad} shows six (1x1) configurations for h-BN within the Ni(111) unit cell. 
The nomenclature of the structure (B,N)=(fcc,top) indicates that boron is sitting on the fcc site, i.e. on a site where no atom is found in the second nickel layer, and where the nitrogen atom sits on top of the atom in the first nickel layer.
Theory found two stable structures (B,N)=(fcc,top) and (hcp,top) only. In both cases nitrogen is on top \cite{gra03}. 
The (fcc,top) structure has the lowest energy, which was consistent with the published experimental structure determinations.
The calculated energy difference between the structure with boron on fcc and on hcp differs only by 9 meV, which is reasonable since it indicates interaction of the overlayer with the second nickel layer.
The 9 meV are, however, one order of magnitude smaller than the thermal energies $k_BT$ during the synthesis, and from this viewpoint it is not clear why pure (B,N)=(fcc,top) single domain can be grown. 
Inspection of Figure \ref{FGrad} shows that the (fcc,top) structure is a translation of the (top,hcp) or (hcp,fcc). 
A transformation between (fcc,top) and (hcp,top) involves, a rotation, or a permutation of B and N and a translation. 
The growth of the h-BN layers on Ni(111) proceeds via the nucleation and growth of triangular islands that are separated by distances between 10 and 100 nm \cite{auw03}.
Therefore, if (fcc,top) and (hcp,top) nucleation seeds form, two domains are expected since the 
change of orientation of islands with sizes larger than 10 nm costs too much energy  \cite{auw03}.
The concomitant domain boundaries (see Figure \ref{intercal} a)) have interesting functionalities.
They act as collectors for intercalating atoms, and clusters like to grow on these lines.
Also, it is expected that such domain boundaries are model systems for $sp^2$ edges.
Unfortunately, the (fcc,top):(hcp,top) ratio is not yet under full experimental control. 
But it is e.g. known that the omission of the oxygen treatment of the Ni(111) surface before h-BN growth causes two domain systems \cite{auw03}.

The vertical position of the nitrogen and the boron atoms is not the same within the (1x1) unit cell. 
This local corrugation or buckling, where nitrogen is the outmost atom, and boron sits closer to the first Ni plane, was taken as an indication of the compressive stress on the h-BN in the h-BN/Ni(111) system \cite{rok97,auw99}. 
The comprehensive density functional theory study of Laskowski et al.  \cite{las08}, indicated however  that this buckling, i.e. height difference between boron and nitrogen persists also for systems with tensile stress in the overlayer, where the substrate lattice constant is larger than that of the h-BN.  
B is closer to the first substrate plane for all investigated cases.
This is a consequence of the bonding to the substrate, where the boron atoms are attracted and the nitrogen atoms are repelled from the surface \cite{las08}. 

For the corresponding graphene Ni system the same structure i.e the (C$_A$,C$_B$)=(fcc,top) which is equivalent to the (top,fcc) configuration has been singled out against the (fcc,hcp) structure \cite{gam97}. 
There are no reports on  (top,fcc)/(top,hcp) domain boundaries, as observed for the h-BN/Ni(111) case \cite{auw03}. 
For g/Ir(111), which belongs to the corrugated layer systems, dislocation lines that are terminated by heptagon-pentagon defects were found \cite {cor08}. Of course these kinds of defects were less likely for BN, since this would imply energetically unfaforable N-N or B-B bonds in the BN network.
%%%%%%%%%%%%%%%%%%%%%%%%%%%%%%%%%%
\subsubsection{Corrugated layers: Moir\'e and dislocation networks }
\label{atomictextured}

When the lattice mismatch ${M}$ of the laterally rigid sp$^2$ networks exceeds a critical value, super structures with large lattice constants are formed. If the lattice of the overlayer and the substrate are rigid and parallel, the super structure lattice constant gets  $a_{ovl}/|{M}|$, where $a_{ovl}$ is the 1x1 lattice constant of graphene or h-BN ($\approx $ 0.25 nm).

\begin{figure}
	\centerline{\includegraphics[width=0.8\textwidth]{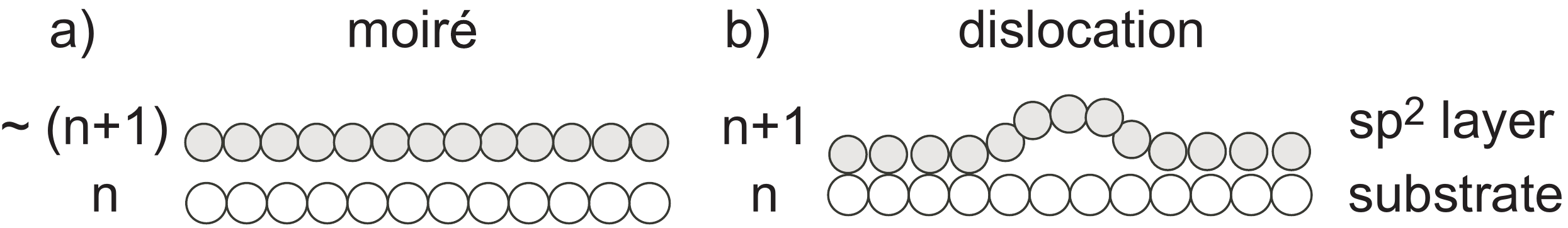}}	
		\caption{Schematic view of a moir\'e and a dislocation in an overlayer system. a) in a moir\'e pattern $\sim (n+1)$ $sp^2$ units fit on $n$ substrate units, and the directions of the substrate and the adsorbate lattice do not necessarily coincide. b) in a dislocation network  $n+1$ $sp^2$ units coincide with $n$ substrate units. Here the lock-in energy is large for on-top and weak or repulsive for bridge sites, where the dislocation evolves.}
		\label{Fdislocation}
\end{figure}

\begin{figure}
	\centerline{\includegraphics[width=0.5\textwidth]{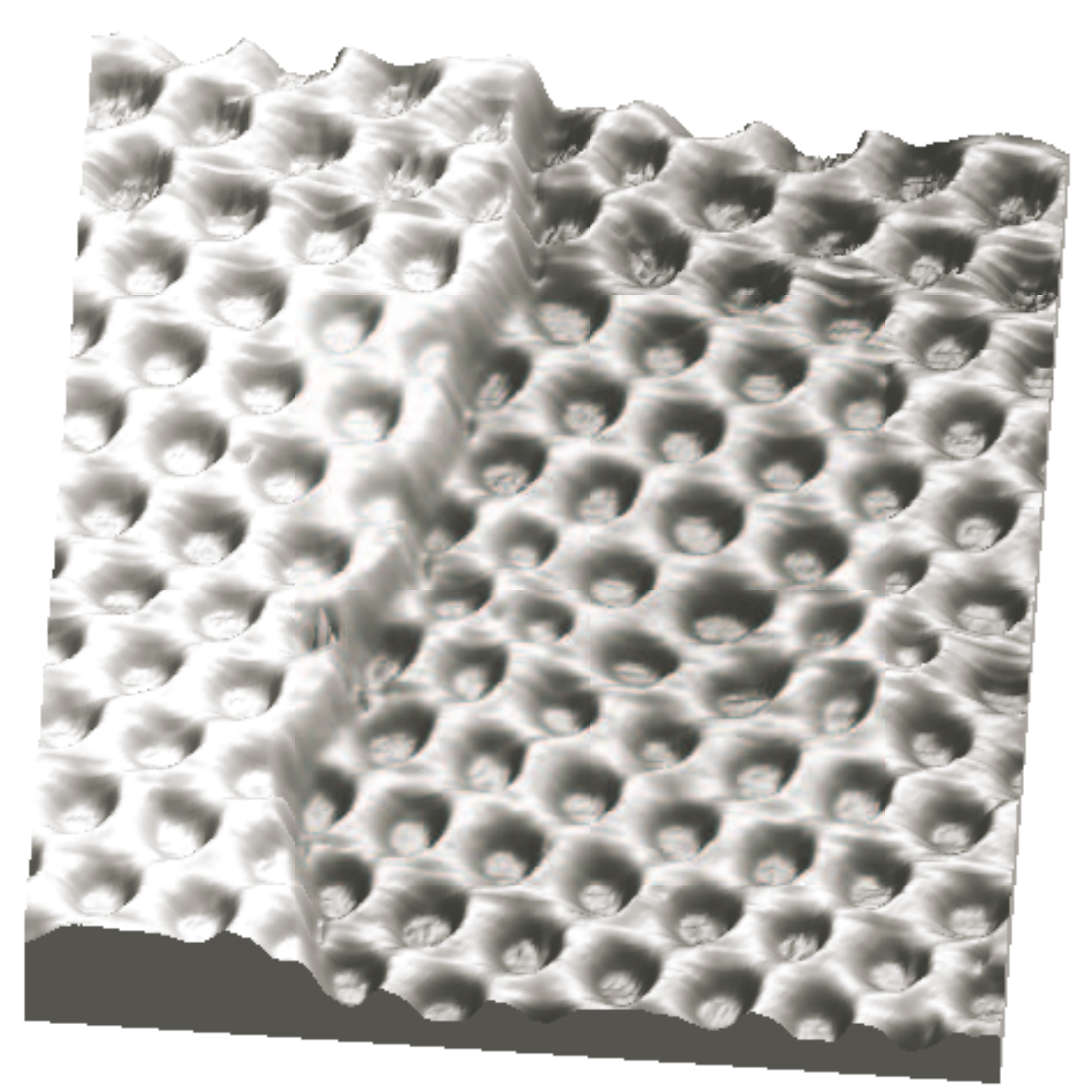}}	
		\caption{Relief view of a constant current scanning tunneling microscopy image of a corrugated layer  boron nitride (nanomesh) (30x30 nm$^2$, I$_t$=2.5 nA, V$_s$=-1 V). This nanostructure with 3.2$\pm$0.1 nm periodicity consists of two distinct areas: the wires which are 1.2+0.2 nm broad and the holes with a diameter of  2.0$\pm$0.2 nm. The corrugation is 0.07$\pm$0.02 nm. From \cite{gre09}.}
		\label{FSTM}
\end{figure}
\begin{figure}
	\centerline{\includegraphics[width=0.4\textwidth]{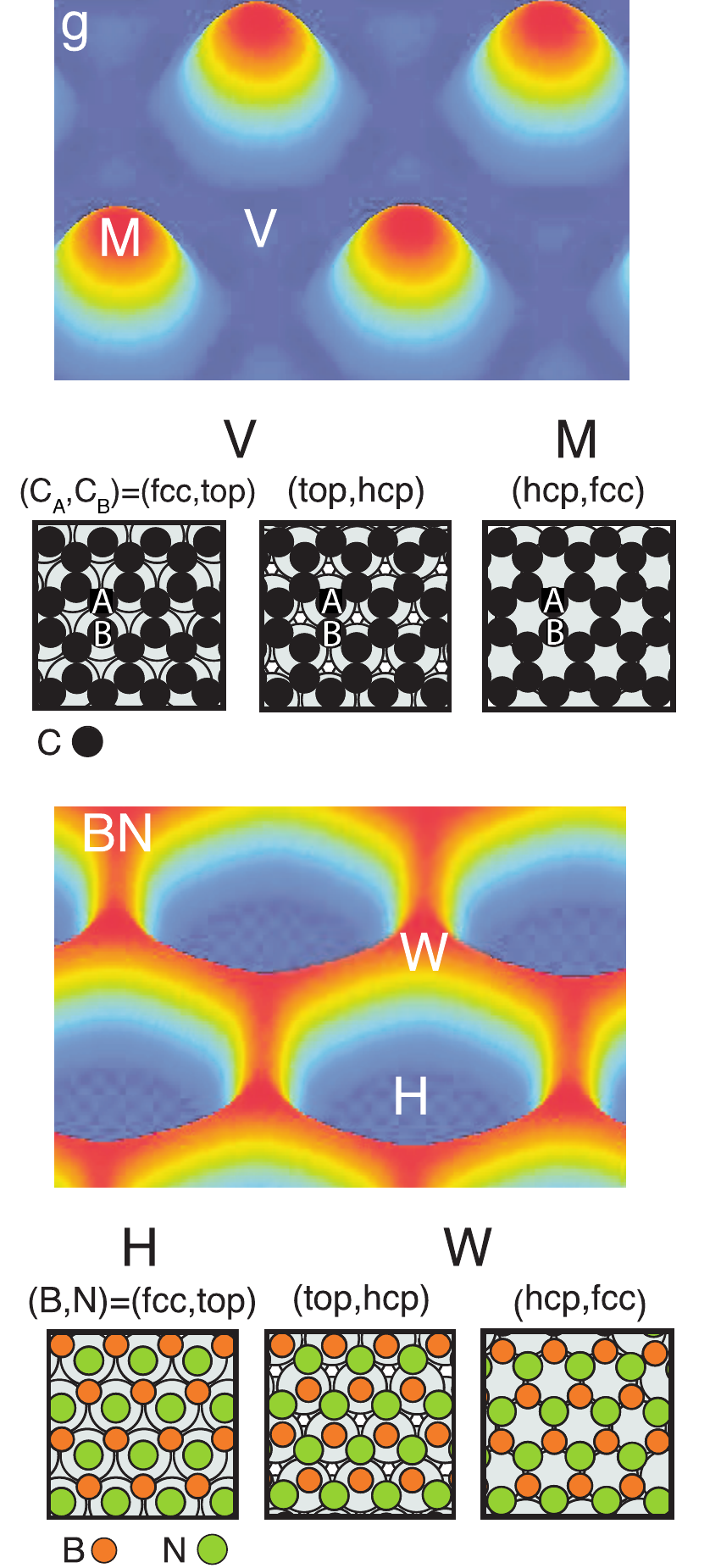}}
	\caption{Views of the height modulated graphene (g) and $h$-BN nanomesh (BN) on Ru(0001), M and V  denote mounds (hcp,fcc) and valleys of the graphene, H and W holes (fcc,top) and wires of the $h$-BN nanomesh. The six ball model panels illustrate the three different regions ((fcc,top), (top,hcp) and (hcp,fcc)), which can be distinguished in both systems \cite{bru09}.}
\label{F1Brugger}
\end{figure}

For rigid $sp^2$ layers, i.e. if there would be no lateral lock-in energy available, we expect flat floating layers, reminiscent to incommensurate moir\'e patterns without a directional lock-in of the structures (see Figure \ref{Fdislocation} a)).
For the case of h-BN/Pd(111) such a tendency to form moir\'e type patterns, also without a preferential lock-in to a substrate direction, were found \cite{mor06}. These h-BN films have the signature of the electronic structure of flat single layers (see Section \ref{es2}).
The lock-in energy is the energy that an epitaxial system gains on top of the average adhesion energy if the overlayer locks into preferred bonding sites.
It involves the formation of commensurate coincidence lattices between the substrate and the adsorbate layer with dislocations (see Figure \ref{Fdislocation} b)).
The lock-in energy is expected to be proportional to the bond energy shown in Figure \ref{FLaskowski}.
For the case of $h$-BN/Rh(111) the lattice mismatch (Eqn. \ref{mismatch}) is -6.7 \%, and
$13\times13$ BN units coincide with $12\times12$ Rh units \cite{cor04,bun07}. 
With the room temperature lattice constants of h-BN and Rh this leads to a residual compression of the 13 h-BN units by 0.9 \%.
Figure \ref{FSTM} shows a relief-view of this super structure coined "h-BN nanomesh" \cite{cor04}.
It has a super cell with a lattice constant of 3.2 nm i.e. (12x12) Rh(111) units and displays the peculiar topography of a mesh with "wires" and "holes" or "pores". 
It turned out that this structure is a corrugated single layer of hexagonal boron nitride with two electronically distinct regions that are related to the topography \cite{las07,ber07}. 
This structure also has the ability of trapping single molecules in its holes (see Section \ref{functextured}).
The accompanying variation of the local coordination of the substrate and the adsorbate atoms divides the unit cells into regions with different registries. In reminiscence to Figure \ref{FGrad} the notation (B,N)=(top,hcp) refers to the local configuration, where a B atom sits on top of the substrate atom in the first layer and N on top of the hexagonal close packed (hcp) site that is on top of the substrate atom in the second layer. Again 3 regions can be distinguished with atoms in (fcc,top), (top,hcp) and (hcp,fcc) configurations (see Figure \ref{F1Brugger}). 
A force field theory approach indicated that the (B,N)=(fcc,top) sites correspond to the tightly bound regions of the h-BN layers, i.e. the holes, while the weakly bound or even repelled regions correspond to the wires \cite{las07}. 
The corrugation, i.e. height difference of the h-BN layer from the top of the substrate is, in accordance between experiment and theory about 0.05 nm.
This is sufficient to produce a distinct functionality (see Section \ref{functextured}) \cite{ber07}.
For h-BN/Ru(0001) a structure very similar to that of h-BN/Rh(111) was found \cite{gor07}.

For the graphene case the situation is related, though not identical.
The difference lies in the fact that the base in the 1x1 unit cell of free-standing graphene consists of two identical carbon atoms C$_A$ and C$_B$, while that of h-BN does not.
C$_A$ and C$_B$ become only distinguishable by the local coordination to the substrate. 
In g/Ru the local (fcc,top) and (top,hcp) coordination leads to close contact between the (C$_{A}$,C$_{B}$) atoms and the substrate \cite{wan08} while (B,N) is strongly interacting only in the (fcc,top) coordination \cite{las08}. As a result, twice as many atoms are bound in strongly interacting regions in g/Ru when compared to $h$-BN/Ru. 
In reminiscence to morphological terms the strongly bound regions of g/Ru(0001) are called  valley (V) and the weakly bound region with the (C$_{A}$,C$_{B}$) atoms on (hcp,fcc) sites mounds (M). The fact that (top,hcp) leads to strong bonding for graphene but weak bonding for $h$-BN gives rise to an inverted topography of the two layers: a connected network of strongly bound regions for graphene (valleys) and a connected network of weakly bound regions for $h$-BN (wires).
Also for the graphene case "moir\'e"-type superstructures were found, where g/Ir(111) is the best studied so far \cite{ndi06}.

%%%%%%%%%%%%%%%%%%%%%%%%
\subsection{Electronic structure I: Work function}
\subsubsection{Flat layers: Vertical polarization}
\label{vertpol}

The work function, i.e. the minimum energy required to remove an electron from a solid, is material dependent.
There are excellent reviews on the topic \cite{hol79,kie81}.
For our purpose, where we want to discuss the electric fields near the surface, it is sufficient to recall the Helmholtz equation that  relates the work function $\Phi$ of a flat surface with vertical electric dipoles:
\begin{equation}
\Phi=\frac{e}{\epsilon_0} N_a \cdot p
\label{Helmholtz}
\end {equation}

where $e$ is the elementary charge, $\epsilon_0$ the permittivity and $N_a$ the areal density of the dipoles $p$.  
If the dipoles are assigned to the atoms, we get e.g. for  Ni(111) with a work function of 
{5.2 eV} and an in plane lattice constant of 0.25 nm  a dipole of 0.7 Debye (1 Debye= 3.34 $\cdot 10^{-30}$ Cm).
It has to be said that this is the classical view of the work function and all quantum mechanical effects may be incorporated empirically into Equation \ref{Helmholtz} in using  an effective dipole.

\begin{figure}
\centerline { \includegraphics[width=0.5\textwidth]{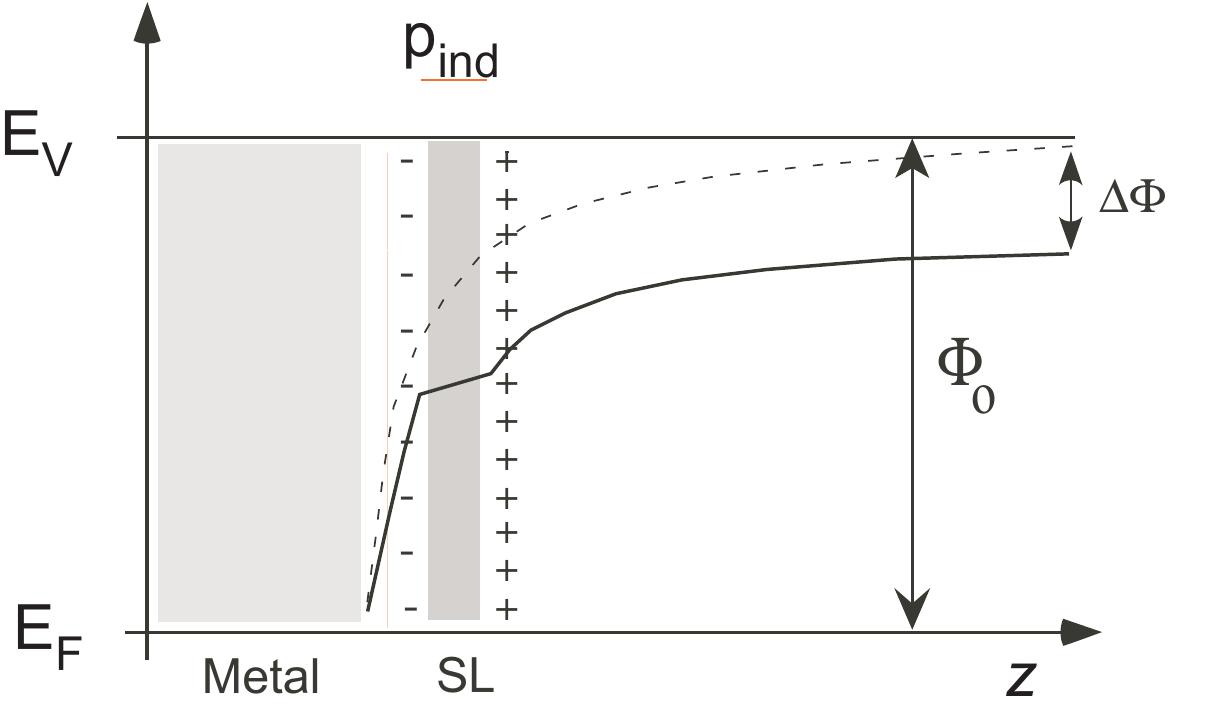}}
\caption{Schematic view of the effect of a thin insulated, polarizable single layer (SL) on the work function. The layer gets polarized and accordingly reduces the surface dipole. The dashed line is the electrostatic potential without single layer, while the solid line shows the effect of the single layer.}
\label{pola}
\end{figure}

Now the influence of overlayers shall be analyzed.
If a single layer of a medium is placed on top of the substrate, the electric fields in the surface dipole polarize the layer i.e. decreases the work function by  $\Delta\Phi_s=\frac{e}{\epsilon_0} N_a \cdot p_{ind}$ by screening.
The induced dipole $p_{ind}$ is proportional to the electric field perpendicular to the surface$E_{\perp}$: $p_{ind}=\alpha \cdot E_{\perp}$, where $\alpha$ is the polarizability.
%The microscopic polarizability is linked via the Clausius-Mosotti equation to the macroscopic dielectric constant. 
%For h-BN we get with $\epsilon=6.85$ $\alpha= 5.72 e^{-40}$
In Figure \ref{pola} it is shown how a polarizable medium screens out the electric field and decreases the work function.
These considerations apply for  dielectrica and metals, if they are not in contact with the substrate.
The strong vertical distance dependence of the electric field in the surface dipole layer involves a correlation with the screening induced work function shift $\Delta\Phi_s$.
In the case of contact of a metallic overlayer with the Fermi level of the substrate, no vertical dipoles in the sense of Figure \ref{pola} are induced, although the work function may change. However, as we know from the Smoluchowski effect \cite{smo41} a corrugation induces a non-uniform surface charge density and thus lateral electric fields (see next Section).

For a dielectric, as it is h-BN, the screening causes, e.g. in the case of h-BN/Ni(111) a work function lowering of 1.6 eV \cite{gra03}, which corresponds to an induced dipole of {0.3 Debye}.
In this case charge gets displaced, but not transferred from the $sp^2$ layer to the substrate.
If the medium is metallic, as it is for most graphene cases, we expect a charge transfer that aligns the chemical potentials of the substrate and the overlayer. 
A strong interaction also alters the chemical potentials and in turn influences the charge transfer. 
In any case the charge redistribution changes the work function, i.e. the surface dipole.
For weakly bound graphene theory predicts a correlation between the work function of the substrate and the doping level \cite{gio08}.
%%%%%%%%%%%%%%%%%%%%%%%%%%%%
\subsubsection{Corrugated layers: Lateral electric fields, dipole rings}
\label{lateraltextured}

\begin{figure}
\centerline {\includegraphics[width=0.8\textwidth]{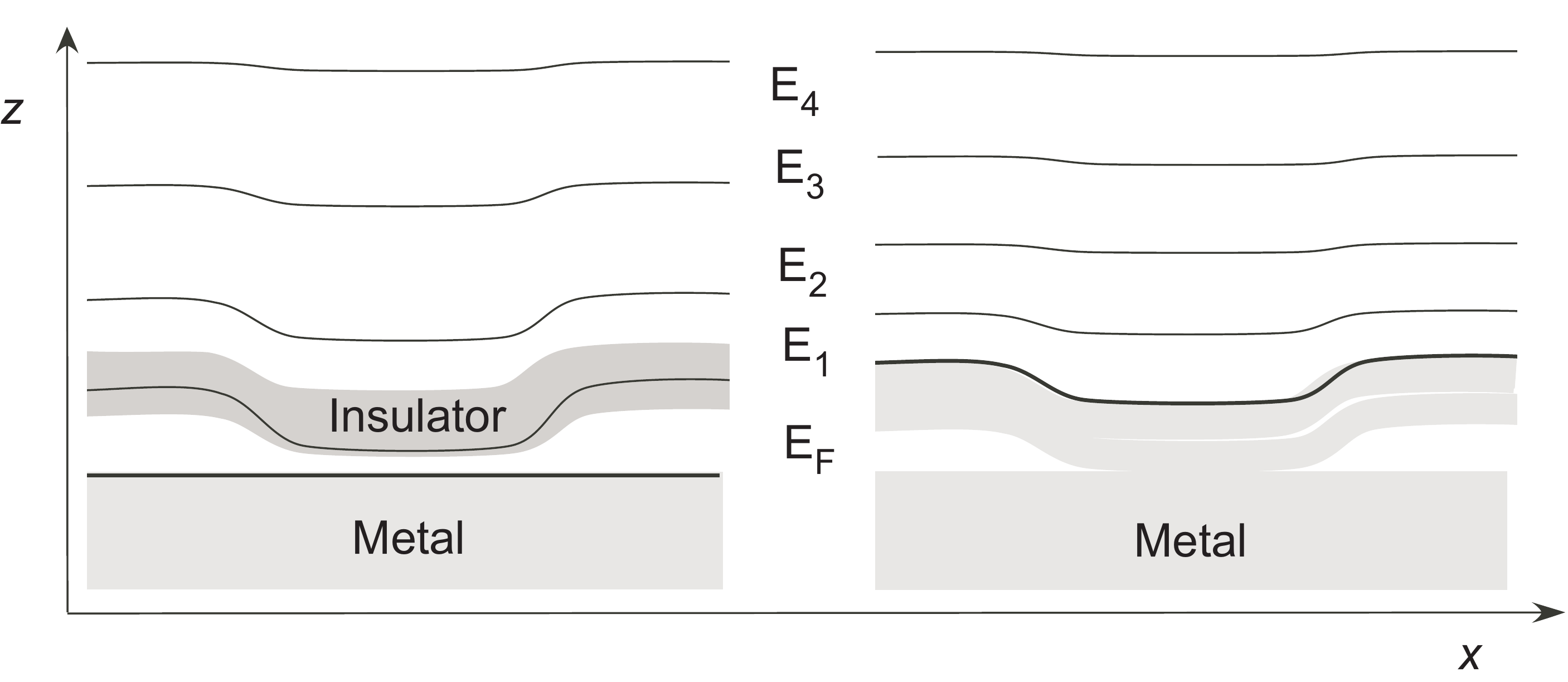}}
\caption{Contour plot of the electrostatic potential at the surface of a corrugated single layer on a flat metal. The energies continuously increase from the Fermi level $E_F$ towards the vacuum level $E_V $ at $z=\infty$, with $E_F<E_1<E_2<E_3<E_4<E_V$. Left for a single layer dielectric, right for a single layer metal.}
\label{potentials}
\end{figure}

If the surface is flat the lateral electric fields will only vary for ionic components in the two sublattices of the $sp^2$ networks.
For the case of h-BN the different electronegativities of the two elements cause a local charge transfer from the boron to the nitrogen atoms. 
By means of density functional theory calculations it has been found that for a free-standing h-BN sheet about 0.56 electrons are transferred from B to N (B,N)=(0.56,-0.56)$e^-$ \cite{gra03}. 
This yields a sizable Madelung contribution to the lattice stability of 2.4 eV per 1x1 unit cell (see Appendix \ref{Elbandstructure}).
If the h-BN sits on a metal this Madelung energy is reduced by a factor of 1/2 due to the screening of the ionic charges.
For h-BN on Ni(111)  the ionicity slightly increases  (B,N)=(0.65,-0.59) $e^-$, where the net charge displacement of 0.06 $e^-$ towards the substrate is consistent with the above mentioned work function decrease due to polarization \cite{gra03}. 
It is expected that the relatively strong local ionicity of the h-BN surface has an influence on the diffusion of atoms with the size of the BN bond length of 0.15 nm.
If the surface is not flat, but corrugated, this causes lateral electric fields on the length scale of the corrugation, which is particularly interesting for trapping atoms or molecules \cite{dil08}.
Lateral fields may occur in dielectric or above metallic overlayers. 
Figure \ref{potentials} shows the lateral and vertical electrostatic potential in a corrugated single layer nanostructure.
For the case of a dielectric (Figure \ref{potentials} a)) a different distance of the layer from the surface imposes a different screening and lateral potential variations, also within the dielectric.
For the case of a metal (Figure \ref{potentials} b)) a corrugation causes  lateral potential variations, reminiscent to the Smoluchowski effect \cite{smo41}, but not within the layer.
Also, it was found for g/Ru(0001) that the metallic case causes smaller lateral potential variations \cite{bru09}.

For dielectrics or insulators the key for the understanding of the lateral electrostatic potential variation came from the $\sigma$ band splitting (see Section \ref{es2textured}).
This splitting, which is in the order of 1 eV,  is also reflected in N1s core level x-ray photoelectron spectra \cite{pre072}, where the peak assignment is in line with the $\sigma$ band assignment \cite{gor07,ber07}.
Without influence of the substrate the energy of the $\sigma$ band, which reflects the in plane $sp^2$ bonds, are referred to the vacuum level. 
This means that the sum of the work function and the binding energy as referred to the Fermi level, is a constant.
Vacuum level alignment arises for physisorbed systems, as e.g. for noble gases \cite{wan84,jan94}, or as it was proposed for h-BN films on transition metals \cite{nag951}.
The $\sigma$ band splitting causes the conceptual problem of aligning the  vacuum level and the Fermi level with two different work functions.
The {\it{local}} work function, or the electrostatic potential near the surface may, however, be different.
This electrostatic potential with respect to the Fermi level can be measured with photoemission of adsorbed Xe (PAX) \cite{kup79,wan84}. 
Xenon does not bond strongly to the substrate and thus the core level binding energies are a measure for the potential energy difference between the site of the Xe core (Xe has a van der Waals radius of about 0.2 nm) and the Fermi level.
Recently, the method of photoemission from adsorbed Xe was successfully applied to explore the electrostatic energy landscape of h-BN/Rh(111)  \cite{dil08}.
In accordance with density functional theory calculations it was found that the electrostatic potential at the Xe cores in the holes of the h-BN/Rh(111) nanomesh is 0.3 eV lower than that on the wires.
This has implications for the functionality, since these sizable potential gradients polarize molecules, and it may be used as an electrostatic trap for molecules or negative ions.
The peculiar electrostatic landscape has been rationalized with dipole rings, where in plane dipoles, sitting on the rim of the corrugations produce the electrostatic potential well.
For in plane dipoles that sit on a ring the electrostatic potential energy in the center of the ring becomes
\begin{equation}
\Delta E_{pot}=\frac{e}{4\pi\epsilon_0}\frac{P}{R^2}
\label{E1}
\end {equation}

\begin{figure}
	\centerline{\includegraphics[width=0.5\textwidth]{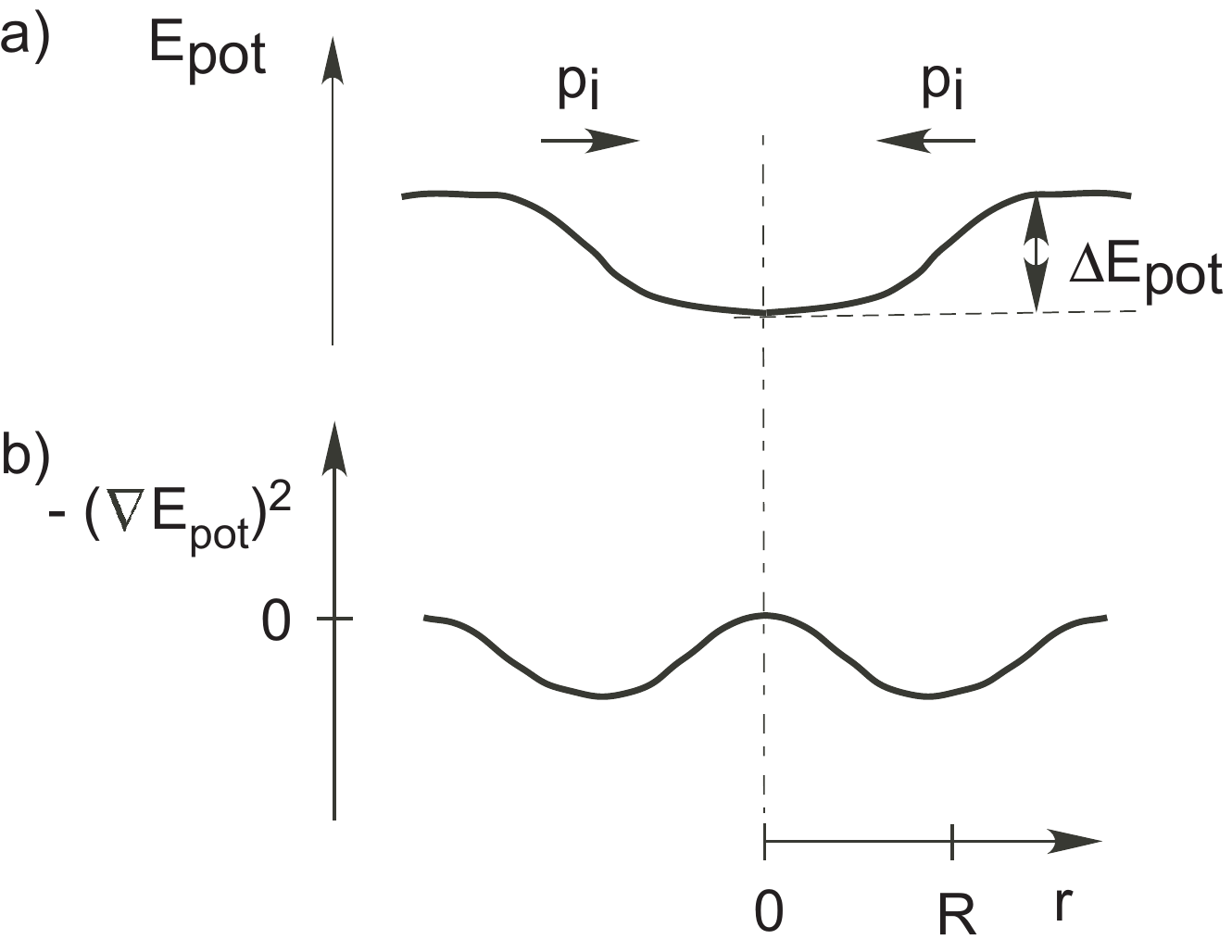}}
		\caption{Schematic drawing of a dipole ring where in plane dipoles ${\bf{p}}_i$ are sitting on a ring with radius $R$. a) the electrostatic potential energy $E_{pot}$ and b) the polarization energy, which is proportional to $(\nabla E_{pot})^2$. From \cite{gre09}.}
		\label{F3}
\end{figure}

where $e$ is the elementary charge, $R$ the radius of the hole and $P=\sum |\textbf{p}_i|$ the sum of the absolute values of the dipoles on the ring. 
For $R$=1 nm and $\Delta E_{pot}$=0.3 eV, $P$ gets 10 Debye, which is equivalent to 5.4 water molecules with the hydrogen atoms pointing to the center of the holes.
The strong lateral electric fields in the BN nanomesh may be exploited for trapping molecules, or negatively charged particles.
It can also act as an array of electrostatic nanolenses for slow charged particles that approach or leave the surface.
Figure \ref{F3} shows the electrostatic potential in a dipole ring and the square of the related electric fields.
The square of the gradient of the electrostatic potential, or the electric field, is expected to be proportional to the polarization induced bond energy  $E_{pol}$:
\begin{equation}
E_{pol}={\bf E}_{\parallel}\cdot {\bf{p}}_{ind}=\alpha \cdot {\bf{E}}_{\parallel}^2
\label{Epol}
\end{equation}
where ${\bf{E}}_{\parallel}$ is the lateral electric field, ${\bf{p}}_{ind}$ the induced dipole, and $\alpha$ the polarizability.
For the case of h-BN/Rh(111) the origin of the in plane electrostatic fields is not the Smoluchowski effect, where the delocalisation of the electrons at steps forms in plane dipoles \cite{smo41}. It is due to the contact of differently bonded boron nitride with different local work functions since the screening depends on the vertical displacement of the dielectric from the metal, i.e. on the corrugation (see Section \ref{vertpol}).

The concept of dipole rings is not restricted to the above mentioned polarization of corrugated dielectrics. 
As it was recently shown for the case of g/Ru(0001),  lateral electric fields also occur above corrugated metals, where the dipoles are created due to lateral polarization like in the Smoluchowski effect \cite{bru09}. 

\subsection{Electronic structure II: Bandstructure, Fermisurfaces}
\label{es2}

The band structure of $sp^2$ layers on transition metals has the same signature as free-standing single layers. The binding energy difference between the in plane $\sigma$ bands and the out of plane $\pi$ bands does not remain constant, when the layers come into contact with a substrate. 
This is a consequence of the bonding via the $\pi$ orbitals, while for the $\sigma$ bands vacuum level alignment is observed.

%%%%%%%%%%%%%%%%%%%%%%%
\subsubsection{Flat layers: The generic case}

The electronic bandstructure of h-BN on Ni(111) is similar to that of a free-standing layer of h-BN, with $\sigma$ and $\pi$ bands that have the generic structure of a $sp^2$ honeycomb lattice (see Appendix). 
In Figure \ref{Fbandstructurehbnni} angle resolved photoemission data from h-BN/Ni(111) along the $\Gamma K $ azimuth are compared with density functional theory calculations.
In the bottom panel the bandstructure calculations are shown for h-BN/Ni(111) and free-standing h-BN. 
The band width of the $\sigma$ bands increases less than 1 \% by the adsorption of the h-BN. The widths of the spin split $\pi$ bands increase in the spin average by 4\%, where the width is larger for the minority spins. 
This band structure with $\sigma$ and $\pi$ bands is generic for all $sp^2$ systems. 
Though, the $\sigma$ and the $\pi$ band do not shift equally in energy upon adsorption of the $sp^2$ layer.
The shift between the $\sigma$ and the $\pi$ bands of about 1.3 eV indicates that the interaction between the Ni substrate and the h-BN $\pi$ and $\sigma$ bands is not the same.
Also other substrates like Pt(111) and Pd(111) cause a similar h-BN band structure, although h-BN/Pt(111) and h-BN/Pd(111) are systems with large negative lattice mismatch \cite{osh97,mor06}.
For these 'flat' cases the bonding to the substrates is rather independent of the positions within the unit cell. In the Section on corrugated layers  (\ref{es2textured}) we will see that this changes when substrates with mismatch and stronger lock-in energy are used.
For Ni(111), Pd(111) and Pt(111) it was found empirically that the $\sigma$ bands have, within $\pm$ 100 meV, the same binding energy, if they are referred to the vacuum level \cite{nag951}. This vacuum alignment confirms that the $\sigma$ bands do not strongly interact with the underlying metal. The binding energy of the $\pi$ band, on the other hand, varies, and indicates the $\pi$ bonding to the substrate. 

For graphene the electronic structure is also well understood \cite{osh97}. 
The $\pi$ bands are of  particular interest since they are decisive for the issue on whether the overlayer is metallic or not.
On substrates the ideal case of free-standing graphene may generally not be realized, since the site selective $p_z$ interaction with the substrate is causing a symmetry breaking between the two carbon sublattices $C_A$ and $C_B$.
This opens a gap at the Dirac point.
The Dirac point coincides with the  $K$ point of the Brillouin zone of graphene, where  the bandstructure is described by cones on which electrons behave quasi relativistic (see Appendix).

\begin{figure}
\centerline {\includegraphics[width=0.65\textwidth]{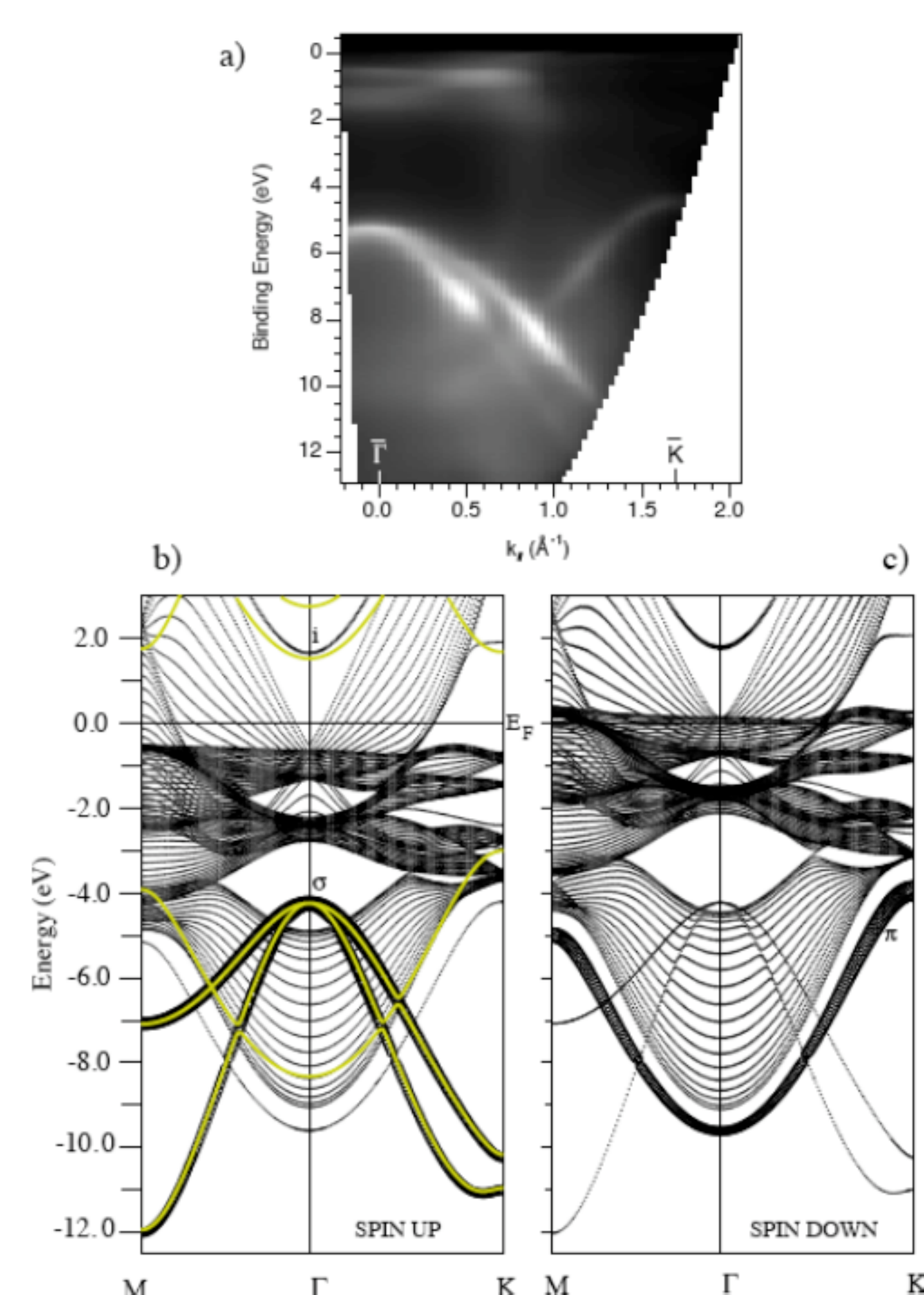}}
\caption{Experimental gray scale dispersion plot on the
$\Gamma K$ azimuth from angle resolved He I$\alpha$ photoemission. The gray scale reproduces the intensity (white
maximum). Bottom: Spin resolved theoretical band structure
of h-BN/Ni(111) along $M \Gamma$ and $\Gamma K$ in the hexagonal reciprocal
unit cell (Brillouin zone). b) Spin up, the radius of the circles is proportional to
the partial $p_x = p_y$ charge density centered on the nitrogen sites ($\sigma$
bands). Thick yellow lines show the bands of a buckled
free-standing h-BN monolayer, which have been aligned the $h$-BN/Ni(111) $\sigma$
band at $\Gamma$. c) Spin down, the radius of the circles is proportional to
the partial $p_z$ charge density on the nitrogen sites ($\pi$ bands). From \cite{gra03}.}
\label{Fbandstructurehbnni}
\end{figure}

\begin{figure}
\centerline {\includegraphics[width=0.5\textwidth]{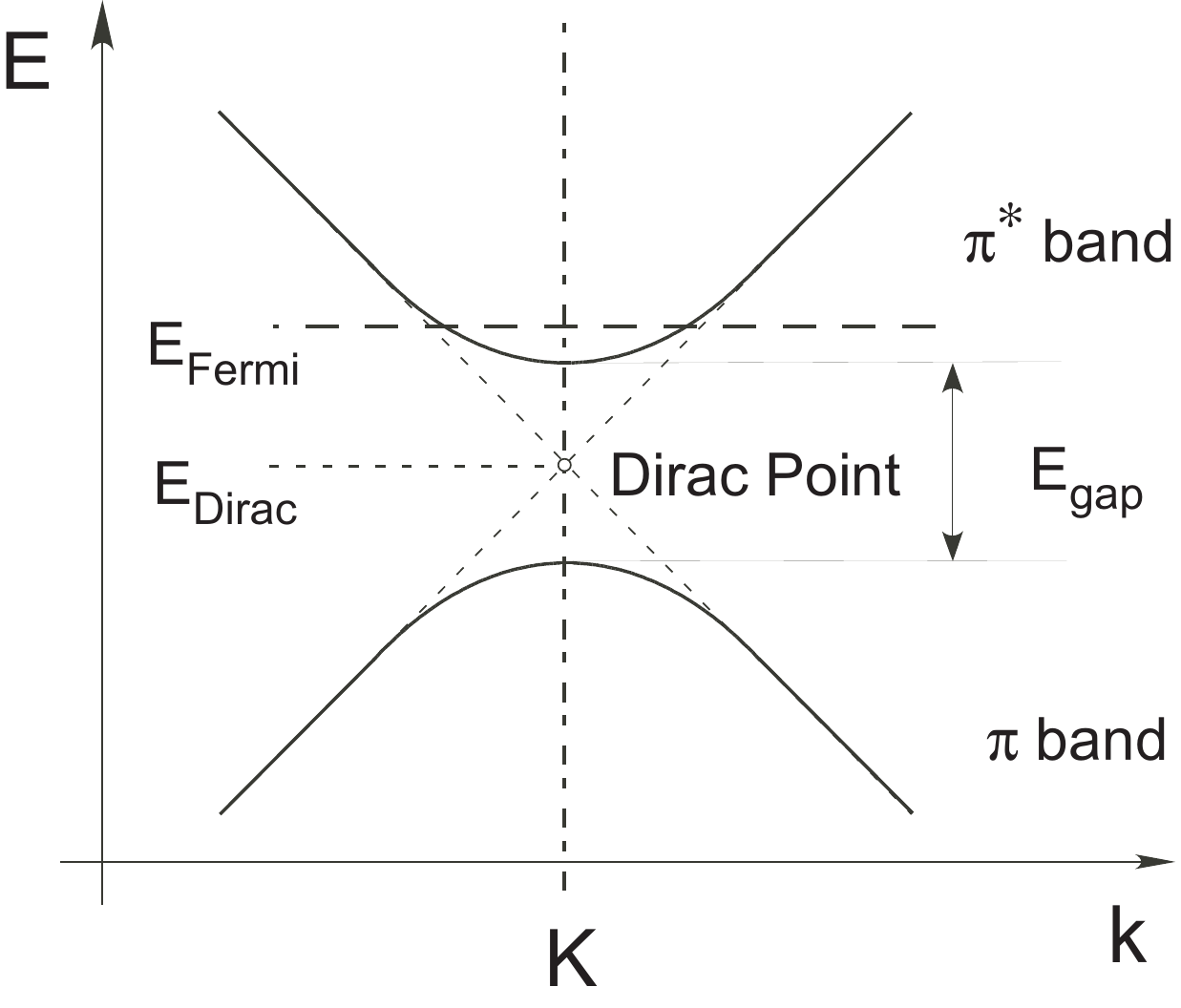}}
\caption{Schematic diagram showing the $\pi$ band electronic structure of $sp^2$ honeycomb lattices around the $K$ point. 
For free-standing graphene the $\pi$ and the $\pi^*$ band lie on cones intersecting at the Dirac point. 
Interaction with a support causes a shift of the Fermi energy, and, if it is anisotropic a gap opens. Here ($E_{Fermi}>E_{Dirac}$), the case for $n$-type conduction is shown (for details see text). }
\label{FDiraccone}
\end{figure}

The magnitude of the gap is a measure for the anisotropy of the interaction of the $C_A$ and the $C_B$ carbon atoms with the substrate. 
The position of the Fermi level with respect to the Dirac energy is decisive on whether we deal with $p$-type or $n$-type graphene (or h-BN).
Since the center of the gap of h-BN lies below the Fermi level, this means that e.g. h-BN/Ni(111) is $n$-type.
Figure \ref{FDiraccone} shows the $\pi$ band electronic structure of $sp^2$ honeycomb lattices near the Fermi energy and around the $K$ point. 
If the Fermi energy lies above the Dirac energy $E_{Dirac}$, which is the energy of the center of the gap, the majority of the charge carriers move like electrons ($n$-type).
Correspondingly, if the Fermi energy lies below the Dirac point the charge carriers move like holes ($p$-type). 
This $n$-type, $p$-type picture is analogous to the semiconductors, where the Fermi level takes, depending on its position, the role of the acceptors ($p$-type) and the donors ($n$-type), respectively.
Nagashima et al. first measured the $\pi$ gaps at $K$ for the case of one and two layers of graphene on TaC(111), where they found  $n$-type behavior and a gap of 1.3 eV for the single layer and about 0.3$\pm$0.1 eV for the double layer \cite{nag94}.    

%%%%%%%%%%%%%%%%%%%%%%%%%%5%
\subsubsection{Corrugated layers: $\sigma$ band splitting}
\label{es2textured}

\begin{figure}
\centerline {\includegraphics[width=0.7\textwidth]{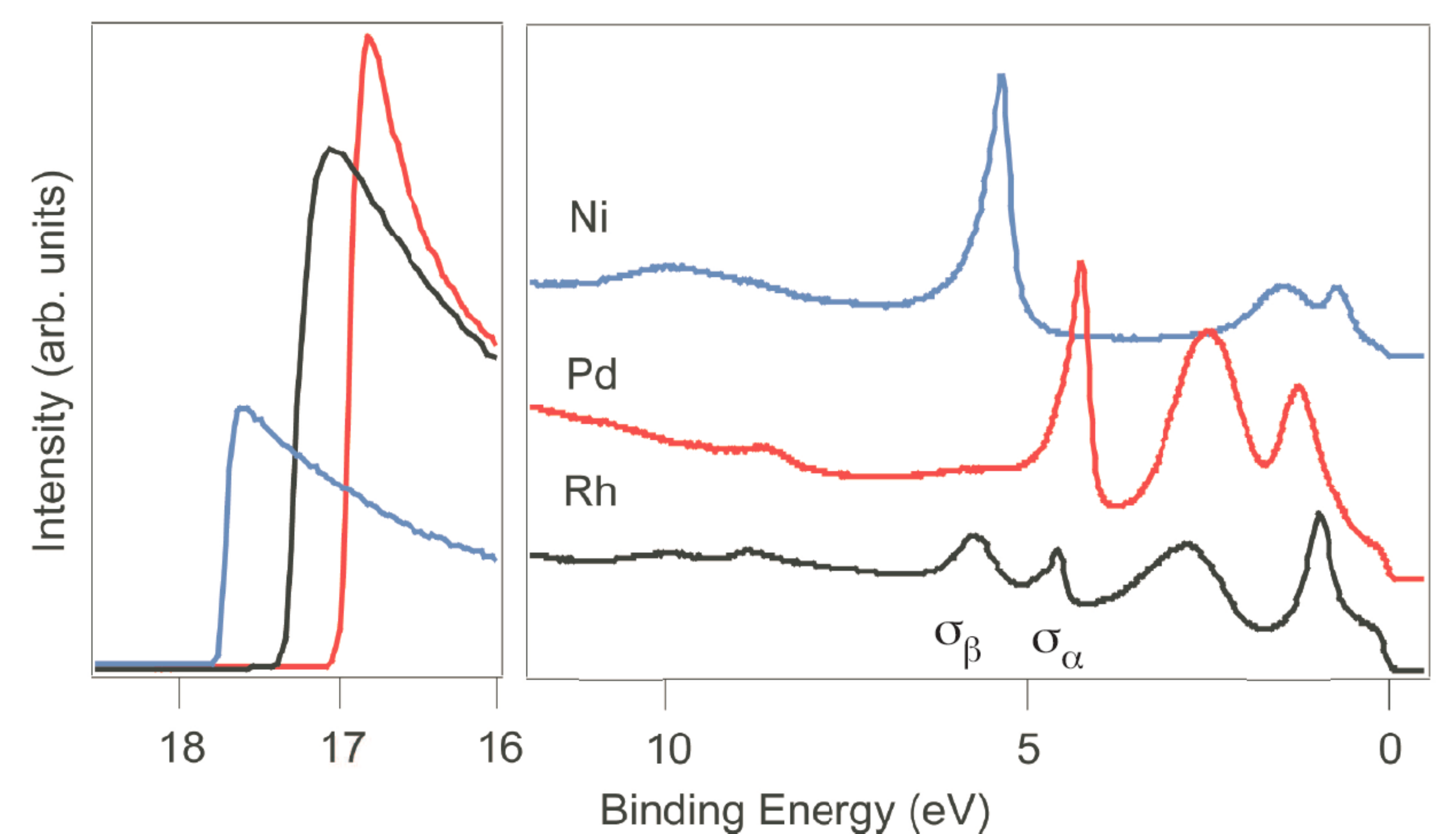}}
\caption{He $I_\alpha$ normal emission photoemission spectra of h-BN/Ni(111), h-BN/Pd(111) and h-BN/Rh(111). While h-BN/Ni(111) and h-BN/Pd(111) show no sizable $\sigma$ band splitting, the splitting into $\sigma_\alpha$ and $\sigma_\beta$ for h-BN/Rh(111) is about 1ÊeV (From \cite{gre09}).}
		\label{F2}
\end{figure}

The super structures as described in Section \ref{atomictextured} are also reflected in the electronic structure.   
Figure \ref{F2} shows valence band photoemission spectra for three different $h$-BN single layer systems in normal emission.
We would like to draw the attention on the $\sigma$ bands at about 5 eV binding energy (see Table \ref{T1}).
Figure \ref{F2} shows valence band photoemission data for h-BN/Ni(111), h-BN/Pd(111) and h-BN/Rh(111). 
Hexagonal boron nitride has 3 $\sigma$ bands and one $\pi$ band that are occupied.
Along ${\Gamma}$, i.e. perpendicular to the $sp^2$ plane, the two low binding energy $\sigma$ bands are degenerate, while the third is not accessible for {He I$_\alpha$} radiation.
Accordingly the $\sigma$ bands of h-BN Ni(111) and of h-BN/Pd(111) show one single peak. 
Those of h-BN/Rh(111) are weak and split in a $\sigma_\alpha$ and a $\sigma_\beta$ contribution \cite{gor07}. 
In measuring the spectra away from the ${\Gamma}$ point it is seen that both, the $\sigma_\alpha$ and the $\sigma_\beta$ bands split in two components each, as it is known from angular resolved measurements of h-BN single layers with $\sigma_\alpha$ bands only \cite{nag951} (see Figure \ref{Fbandstructurehbnni}).

\begin{table}
\caption{Experimental values (in eV) for the photoemission binding energies referred to the Fermi level $E_B^F$ of the $\sigma$ bands for h-BN single layers on three different substrates along $\Gamma$, and the corresponding work functions $\Phi$. The binding energies with respect to the vacuum level $E_B^V$ are determined by $E_B^F+\Phi$.}
 \begin{tabular}{c|c|c|c|c|c}
 \\
 \hline
 Substrate&$E_B^FÊ\sigma_\alpha$&$E_B^FÊ\sigma_\beta$&$\Phi$&$E_B^VÊ\sigma_\alpha$&Reference\\
 \hline
 Ni(111)&5.3& &3.5&8.8&\cite{gre02}\\
 Rh(111)&4.57&5.70&4.15&8.7&\cite{gor07}\\
 Pd(111)&4.61&&4.26&8.9&\cite{gre09}\\
  \hline
 \end{tabular}
\label{T1}
\end{table}

The  $\sigma$ band splitting indicates two electronically different regions within the h-BN/Rh(111) unit cell.
They are related to strongly bound and weakly bound h-BN. 
Later on, h-BN/Ru(0001) \cite{gor07}  was found to be very similar to h-BN/Rh(111).  Theoretical efforts \cite{las07} and atomically resolved low temperature STM \cite{ber07} showed that the h-BN/Rh(111) nanomesh is a corrugated single sheet of h-BN on Rh(111) (see Section \ref{atomictextured}).
The peculiar structure arises from the site dependence of the interaction with the substrate atoms that causes the corrugation of the h-BN sheet, with a height difference between strongly bound regions and weakly bound regions of about 0.05Ênm. 

\begin{figure}
\centerline {\includegraphics[width=0.8\textwidth]{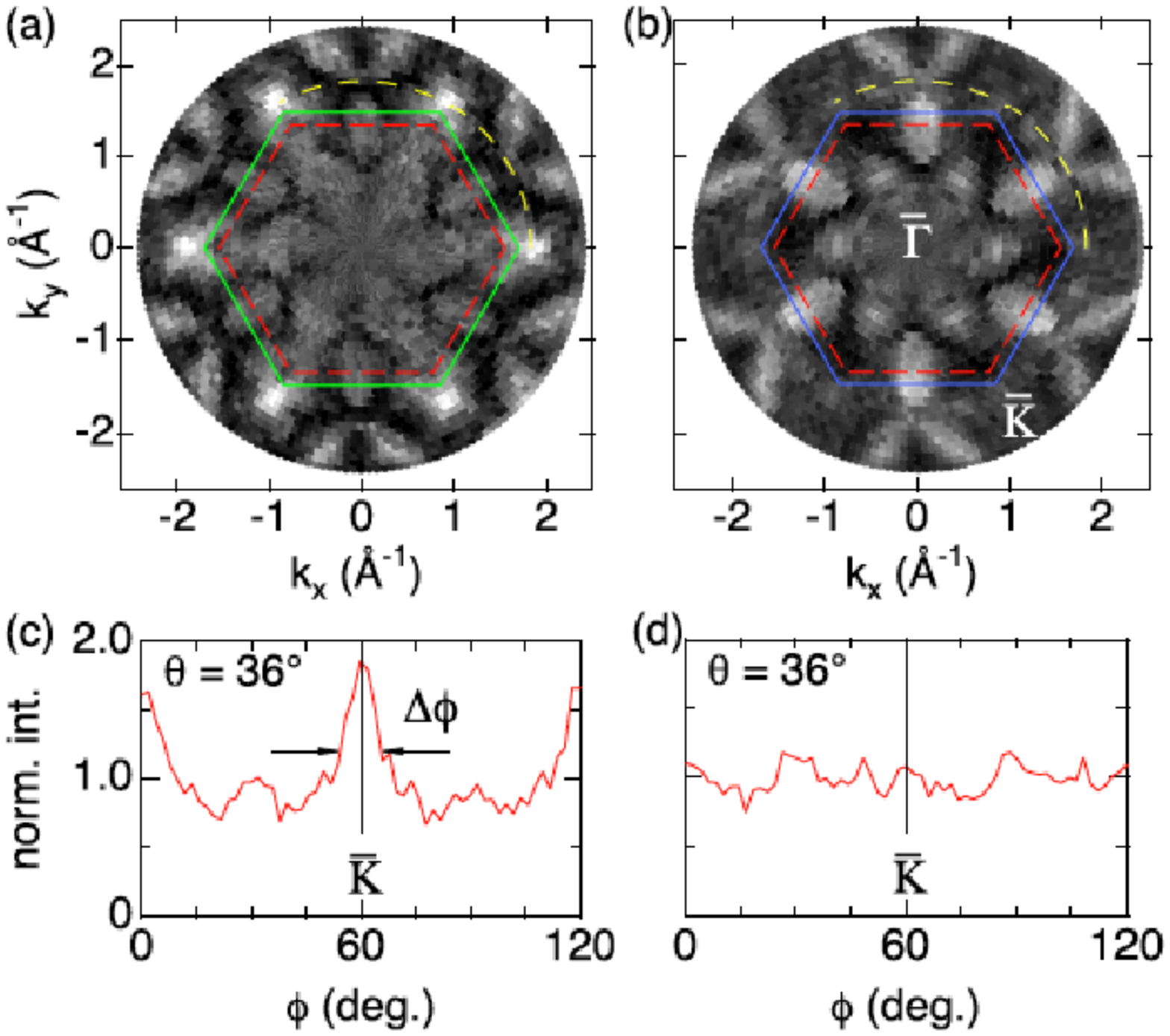}}
\caption{Fermi surface maps. (a) g/Ru(0001). (b) $h$-BN/Ru(0001). The hexagons indicate the surface Brillouin zones of Ru(0001) (red dashed), graphite (green solid) and $h$-BN (blue solid). (c) and (d) show the normalized intensities of azimuthal cuts along the dashed yellow sectors in (a) and (b) respectively. From \cite{bru09}.}
\label{FFSM}
\end{figure}

Interestingly, no $\sigma$ band splitting could be found for the g/Ru(0001) system, although these graphene layers are also strongly corrugated \cite{bru09}. 
This is likely related to the inverted topography (see Figure \ref{F1Brugger}) and the metallicity of graphene on ruthenium. 
The metallicity can be investigated locally by means of scanning tunneling spectroscopy.
It has however to be said that tunneling spectroscopy is most sensitive to the density of states at the $\Gamma$ point but the graphene conduction electrons reside around the  $K$ point. 
This experimental limitation does not exist in angular resolved photoemission, where on the other hand the sub-nanometer resolution of scanning probes is lost. 
With photoemission it is possible to determine the average Fermi surface of the $sp^2$ layers. 
The Fermi surface indicates the doping level, i.e. the number of electrons in the conduction band. 
In the case of graphene cuts across Dirac cones (see Figure \ref{FDiraccone}) are measured.
If the Fermi surface map is compared to another constant energy surface near the Fermi energy, it can be decided whether the Dirac point lies below or above the Fermi energy, i.e. whether we deal with $n$- or $p$-type graphene.
In Figure \ref{FFSM} the Fermi surface map as measured for g/Ru(0001) and h-BN/Ru(0001) are compared.
For the case of g/Ru(0001) extra intensity at the $K$ points indicates Dirac cones. 
The cross sections of which correspond to 5\% of the area of the Brillouin zone, or 0.1 $e^-$.
The cross section shown in Figure \ref{FFSM} c) does not resolve a single cone with a bimodal distribution around $K$, as it is e.g. the case for g/SiC \cite{oht06}, but a single peak from which an average cone diameter at the Fermi energy is determined.
The measurement of the bandstructure along $\Gamma K$ indicates that these 0.1 $e^-$ are donated from the substrate to the graphene.
For the case of h-BN/Ru(0001) the Fermi surface shows no peak at $K$ which is expected for an insulating or semiconducting  $sp^2$ layer \cite{gre09}. 
All features that can be seen in this Fermi surface map stem from the underlying Ru(0001) substrate.

%%%%%%%%%%%%%%%%%%%%
%%%%%%%%%%%%%%%%%%%%
\subsection{Sticking and intercalation}
\label{intercalation}

Graphite is well known for its ability to intercalate atoms \cite{dre81}, and there are also reports on intercalation in h-BN \cite{bud04}.
Intercalation is the reversible inclusion of guest atoms or molecules between other host molecules.
For the host material graphite or $h$-BN intercalation occurs between the honeycomb sheets,
where the bonding in the sheet is strong and the bonding between the sheets is relatively weak.
Here, we deal with single $sp^2$ layers on transition metals and use the term "intercalation" also for irreversible intercalation, as it is observed if metal atoms slip below the $sp^2$ layer.
For the case of g/Ni(111) it has e.g. been found that Cu, Ag and Au intercalate irreversibly, although they form no graphite intercalation compounds \cite{shi00}.

\begin{figure}
\centerline {\includegraphics[width=0.8\textwidth]{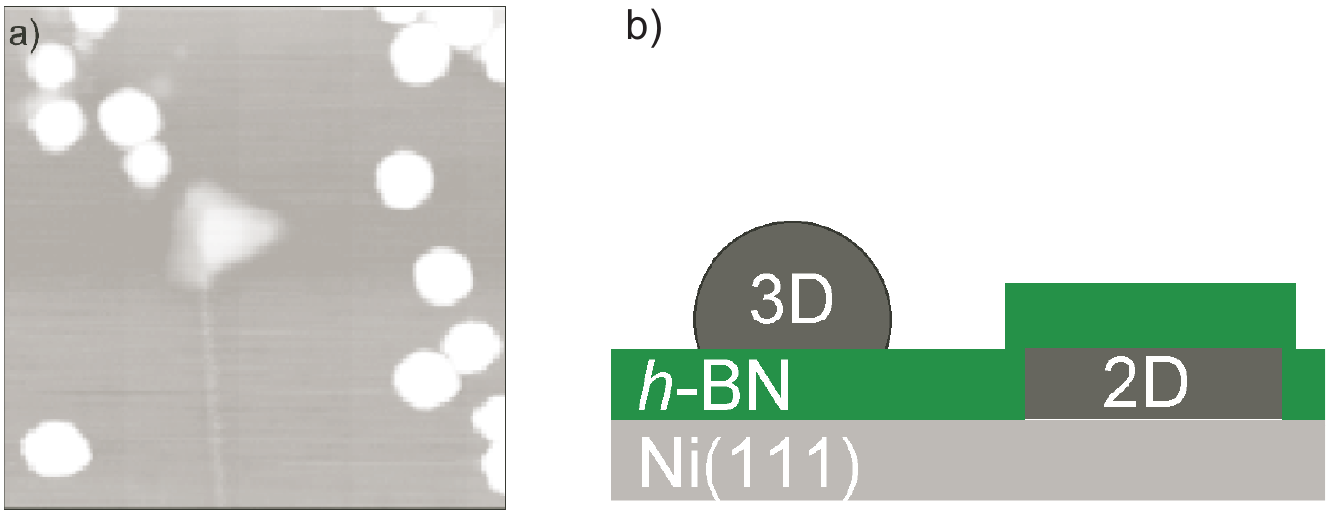}}
\caption{Cobalt on h-BN/Ni(111). a) Scanning tunneling microscopy image, showing triangular two dimensional (2D) intercalated Co islands, and circular three dimensional (3D) clusters and a defect line connecting to a intercalated island (27x27 nm). b) Schematic side view of the situation, showing the topology of the 3D and the 2D agglomerates. Data from \cite{auw02}.}
\label{intercal}
\end{figure}

Intercalation is preceded by sticking (adsorption), and diffusion of the intercalating species. 
In the following the model case of cobalt on h-BN/Ni(111) is discussed in more detail.
Figure \ref{intercal} shows the growth of cobalt on h-BN/Ni(111).
On flat terraces, as shown in Figure \ref{intercal} a) three different patterns are observed. (i) three dimensional (3D) clusters, whose heights scale with the lateral diameter, (ii) triangular, two dimensional (2D) islands with a constant apparent height and (iii) line patterns \cite{auw02}. 
Often these line patterns are found to be connected to 2D islands, and 3D islands tend to nucleate on such lines.
A careful analysis relates the lines with domain boundaries, where (B,N)=(fcc,top) and (B,N)=(hcp,top) domains touch \cite{auw03} (see Section \ref{atomicflat}).
The 2D islands are irreversibly intercalated Co below the h-BN layer, while the 3D clusters remain on top of the h-BN. 
This results e.g. in the property that the 3D islands may be removed, cluster by cluster, with a STM manipulation procedure \cite{auw02}.

\begin{figure}
\centerline{\includegraphics[width=0.8\textwidth]{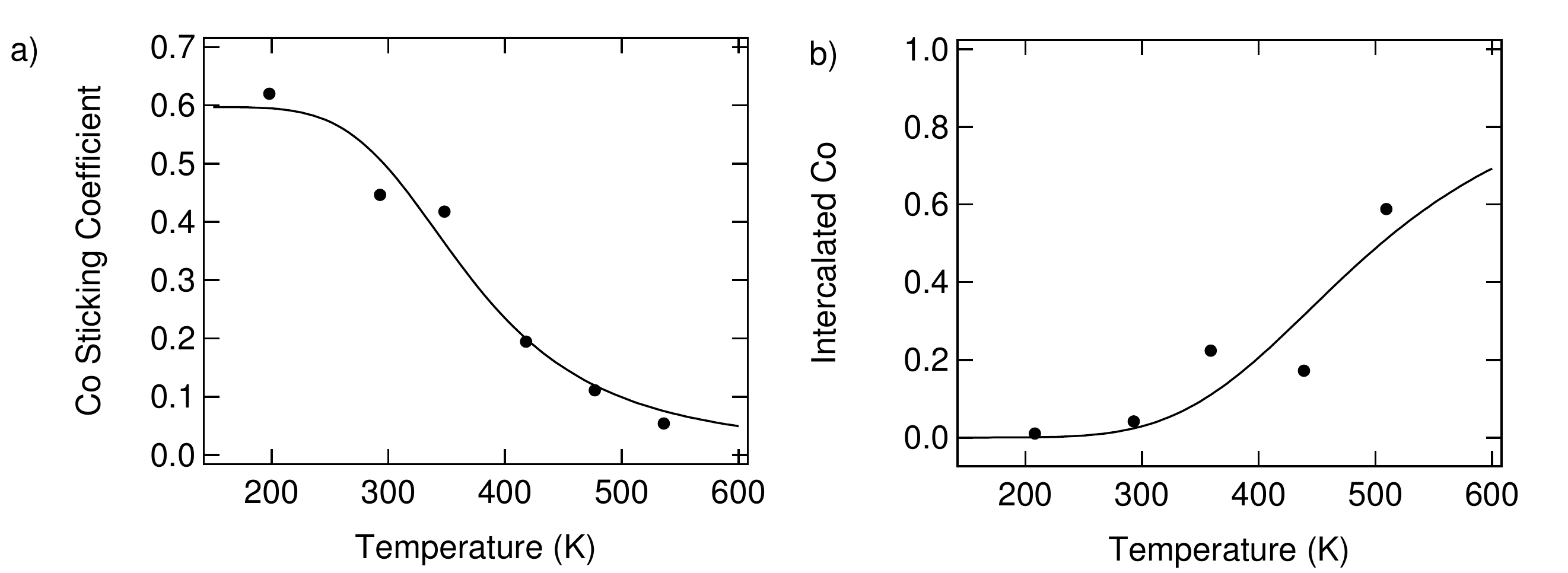}}
\caption{Temperature dependence of a) sticking coefficient of Co from a sublimation source at about 1400 K and b) intercalation of Co below h-BN on Ni(111). Data from \cite{auw02}.}
\label{intercalXPS}
\end{figure}

The sticking coefficient, i.e. the probability that an impinging atom sticks on the surface, is determined by X-ray photoelectron spectroscopy (XPS). While the sample was exposed to a constant flux of about 3.5 monolayers of Co per hour, the measurement of the Co uptake on the sample gives a measure for the sticking probability \cite{auw02}. 
Figure \ref{intercalXPS} a) shows the temperature dependence of the sticking coefficient of cobalt atoms that are evaporated by sublimation onto h-BN/Ni(111).
Clearly, the sticking is not unity, as it is commonly assumed for a metal that is evaporated onto a metal.
Also, the sticking is strongly temperature dependent. 
This means that at higher temperature more Co atoms scatter back into the vacuum and that the bond energy of the individual Co atoms must be fairly small.
The solid line in Figure \ref{intercalXPS} a) indicates the result of an extended Kisliuk model \cite{kis57} for the sticking of Co on h-BN/Ni(111). 
This model predicts about 30\% of the Co atoms not to thermalize on the surface and to directly scatter back, and says that the activation energy for diffusion is about 190 meV  smaller than the desorption energy \cite{auw02}.

The intercalation is also thermally activated.
In Figure \ref{intercalXPS} b) the intercalated amount of Co as compared to the total amount of Co on the surface is shown for low Co coverages.
The fact that at low substrate temperatures almost no Co slips below the $h$-BN, indicates that the intercalation must be thermally activated.
The number  $N^{2D}$ of intercalated atoms, may be modeled with a simple ansatz like:
\begin{equation}
N^{2D}=k_2 \exp{-\frac{E_A^{2D}}{k_BT}}
\end{equation}

where $k_2$ is the reaction constant that comprises the sticking and the diffusion of Co to the intercalation site, $E_A^{2D}$ the thermal activation energy for intercalation at this site and $k_BT$ the thermal energy.
For the 3D clusters the number  $N^{3D}$ of atoms in the clusters is assumed to be constant:
\begin{equation}
N^{3D}=k_3
\end{equation}
where $k_3$ is the reaction constant that comprises the sticking and the diffusion of Co to the 3D cluster nucleation sites.
The backreaction, i.e. the dissolution of atoms from 2D islands or 3D clusters is neglected, since there the Co bond energies are much larger.
The solid line in Figure  \ref{intercalXPS} b) shows a fit with $E_A^{2D} = 0.24 \pm 0.1 \rm{eV}$ and a ratio $\frac{k_2}{k_3}=\frac{1}{250}$.
In $\frac{k_2}{k_3}$ the temperature dependence of the sticking and the diffusion cancel. 
A temperature dependence of the 3D nucleation site density is not included in this kinetic modeling, because the data do not allow the extraction of more than two independent parameters.
The small value of $\frac{k_3}{k_2}$ is an indication that a Co atom finds an intercalation site more easily than a 3D cluster. 
This is consistent with the assignment that the defect lines (see Figure \ref{intercal} a)) act as collectors for the intercalation, since the probability for a diffusing particle to hit a line is much larger than hitting a point like 3D cluster.

%%%%%%%%%%%%
%%%%%%%%%%%%
\subsection{Functionality}

The functionality of $sp^2$ single layers bases on a set of attractive properties like a, compared to clean transition metal surfaces, low reactivity and high thermal stability.
They also feature a change in electron transport perpendicular and parallel to the surface.
In a metal $sp^2$-layer metal hetero-junction the layer changes the phase matching of the electrons in the two electrodes. 
This is particularly interesting in view of spintronic applications where in a magnetic hetero-junction the resistance for the two spin components is not the same, and magnetoresistance or spin filtering is expected \cite{kar07}.
If the layers are insulating or dielectric, as it is the case for h-BN they act as atomically sharp tunneling junctions.
The corrugated $sp^2$ layers bear,  on top of the potential of the flat layers, the possibility of using them as templates for molecular architectures.    

%%%%%%%%%%%%%%%%%%%%%%%%
\subsubsection{Flat layers: Tunneling junctions}
\label{tunnelingbarriers}
 
\begin{figure}
\centerline {\includegraphics[width=1.0\textwidth]{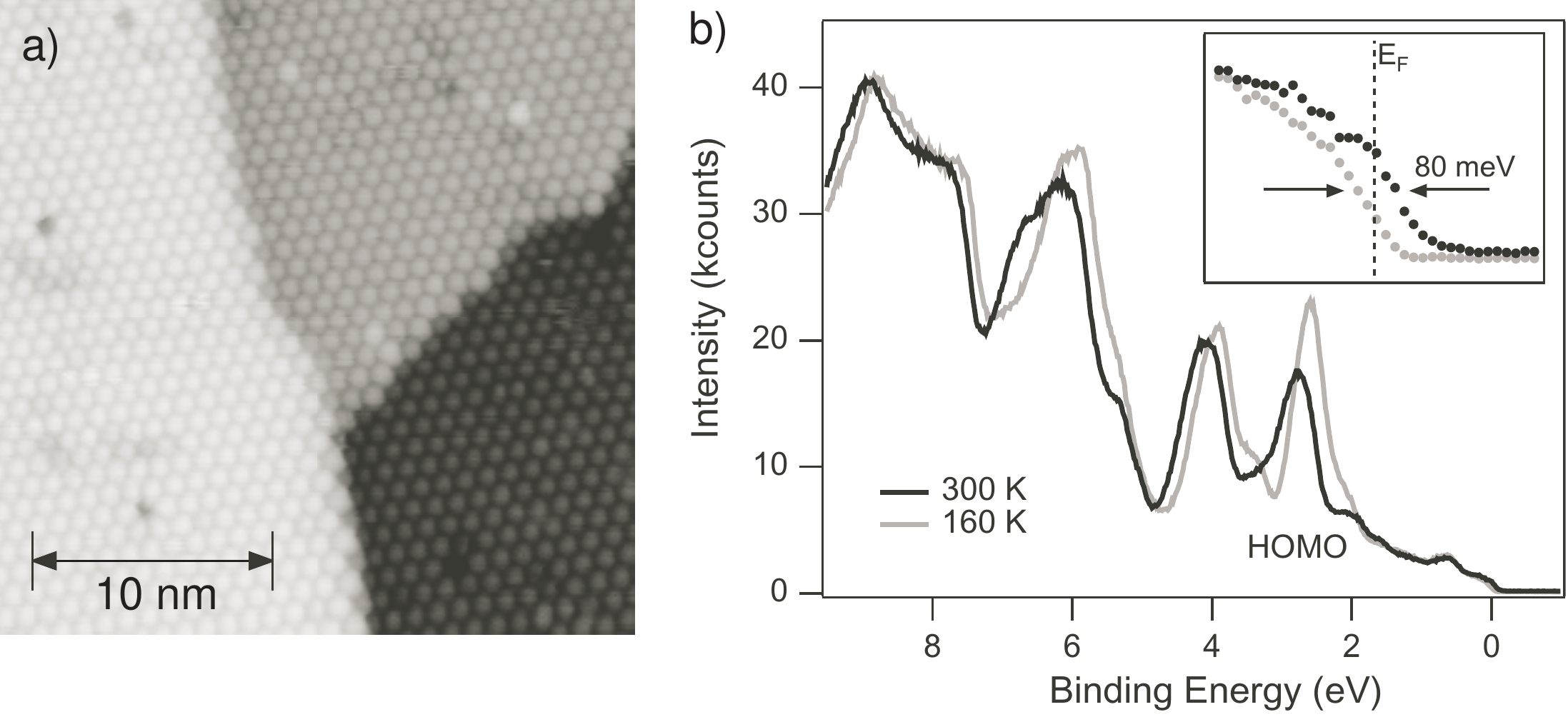}}
\caption{$C_{60}$ on h-BN/Ni(111). a) Scanning tunneling microscopy shows that a monolayer of $C_{60}$ wets the substrate and forms a regular 4x4 structure. b) The valenceband photoemission spectrum is dominated by molecular orbitals of $C_{60}$. The intensity and energy position of the orbitals is strongly temperature dependent. Note the shift of the photoemission leading edge of 80 meV, which corresponds to the transfer of about half an electron onto the molecules in going from 150 K to room temperature. From \cite{mun05}. }
\label{c60hbnni}
\end{figure}

An insulating single layer on a metal acts as a tunneling junction between the substrate and the adsorbate.
It decouples adsorbates from the metal.
This decoupling also influences the time scale on which the electronic system equilibrates.  
The electronic equilibration time scale increases and eventually becomes comparable to that of molecular motion.
This opens the doors for physics beyond the Born-Oppenheimer approximation, where it is assumed that the electronic system is always in equilibrium with the given molecular coordinates. 
Such an example is the C$_{60}$/h-BN/Ni(111) junction \cite{mun05}.
Figure \ref{c60hbnni} a) shows 3 substrate terraces separated by the Ni(111) terrace height of 0.2 nm.  The whole surface is wetted by a hexagonally closed packed C$_{60}$ layer with a 4x4 super structure, with one C$_{60}$ on 4 h-BN/Ni(111) unit cells.
The electronic structure of the C$_{60}$ layer shows a distinct temperature dependence that is in line with the onset of molecular libration. 
The freezing of the molecular rotation at low temperature is well known from C$_{60}$ in the bulk of fullerene \cite{hei91} and  C$_{60}$ at the surface of fullerene \cite{gol96}.
The C$_{60}$ molecular orbitals as measured with photoemission shift by about 200 meV in going from 100 K to room temperature. 
The shift is parallel to the work function, which indicates a vacuum level alignment of the  molecular orbitals of C$_{60}$.
This shift signals a significant charge transfer from the substrate to the adsorbate.
As can be seen in the inset of Figure \ref{c60hbnni} b) it causes an upshift of the leading edge in photoemission. 
This upshift translates in an average charge transfer of about 0.5 $e^-$ onto the lowest unoccupied molecular orbital (LUMO) of the C$_{60}$ cage.
The LUMO occupancy therefore changes by about a factor of 7 in going from 100 K to room temperature, which is a molecular switch function.
The normal emission photoemission intensity of the highest occupied molecular orbital (HOMO) as a function of the substrate temperature shows the phase transition between 100 K and room temperature. 
The fact that the phase transition is visible in the normal emission intensity gives a direct hint that the molecular motion that leads to the phase transition must be of rocking type (cartwheel)  since azimuthal rotation of the C$_{60}$ molecules would not alter this intensity.
The experiments also show that C$_{60}$ is indeed weakly bound to the substrate, since it desorbs at temperatures of $\sim$500 K, which is only 9\% higher than the desorption temperature from bulk C$_{60}$.
If we translate the work function shift in the phase transition into a pyroelectric coefficient $p_i=\partial{P_S}/\partial{T}$, where $P_S$ is the polarization we get an extraordinary high value of -2000 $\mu$C/m$^2$K, which is larger than that of the best bulk materials.

The correlation of the molecular orientation with the charge on the LUMO can be rationalized with the shape of the LUMO of the C$_{60}$. 
The LUMO wave function is localized on the pentagons of the  C$_{60}$ cage and at low temperature  C$_{60}$ does not expose the pentagons towards the h-BN/Ni(111).
Therefore the charge transfer is triggered by the onset of rocking motion where the LUMO orbitals get a larger overlap with the Fermi sea of the nickel metal. 
In turn this will increase electron tunneling to the LUMO.
In an adiabatic picture, however, the back tunneling probability is as large as the forth tunneling probability, and the magnitude of the charge transfer effect could not be explained.
It was therefore argued that the magnitude of the effect is a hint for non adiabatic processes, i.e. that the back tunneling rate gets lower due to electron self trapping.
This self trapping can only be efficient, if the electron tunneling rate is low, i.e. if the electron resides for times that the molecule needs to change its coordinates.

%%%%%%%%%%%%%%%%%%%%%%
\subsubsection{Corrugated layers: Templates}
\label{functextured}

Templates are structures that are able to host objects in a regular way.
In nanoscience they play a key role and act as a scaffold or construction lot for supramolecular self assembly that allow massive parallel production processes on the nanometer scale.
Therefore the understanding of the template function is of paramount importance.
It requires the exploration of the atomic structure that defines the template unit cell geometry, and the electronic structure that imposes the bond energy landscape for the host atoms or molecules.
On surfaces the bond energy landscape has a deep valley perpendicular to the surface that is periodically corrugated parallel to the surface. 
The perpendicular valley is responsible for the adsorption and the parallel corrugation governs surface diffusion.
If the surface shall act as a template, and e.g. impose a lateral ordering on a length scale larger than the (1x1) unit cell, it has to rely on reconstruction and the formation of super structures. 
For the case of the h-BN/Rh(111) the superstructure has a size of 12x12 Rh(111) unit cells on top of which 13x13 h-BN unit cells coincide (see Section \ref{atomictextured}).
It is a corrugated $sp^2$ layer which displays a particular template function for molecular objects with a size that corresponds with the nanostructure.
The super cell divides into different regions, the wires, where the layer is weakly bound to the substrate and the "holes" or "pores" where h-BN is tightly bound to the substrate. 
The holes have a diameter of 2 nm and are separated by the lattice constant of 3.2 nm (see Section \ref{atomictextured}).

\begin{figure}
\centerline {\includegraphics[width=0.9\textwidth]{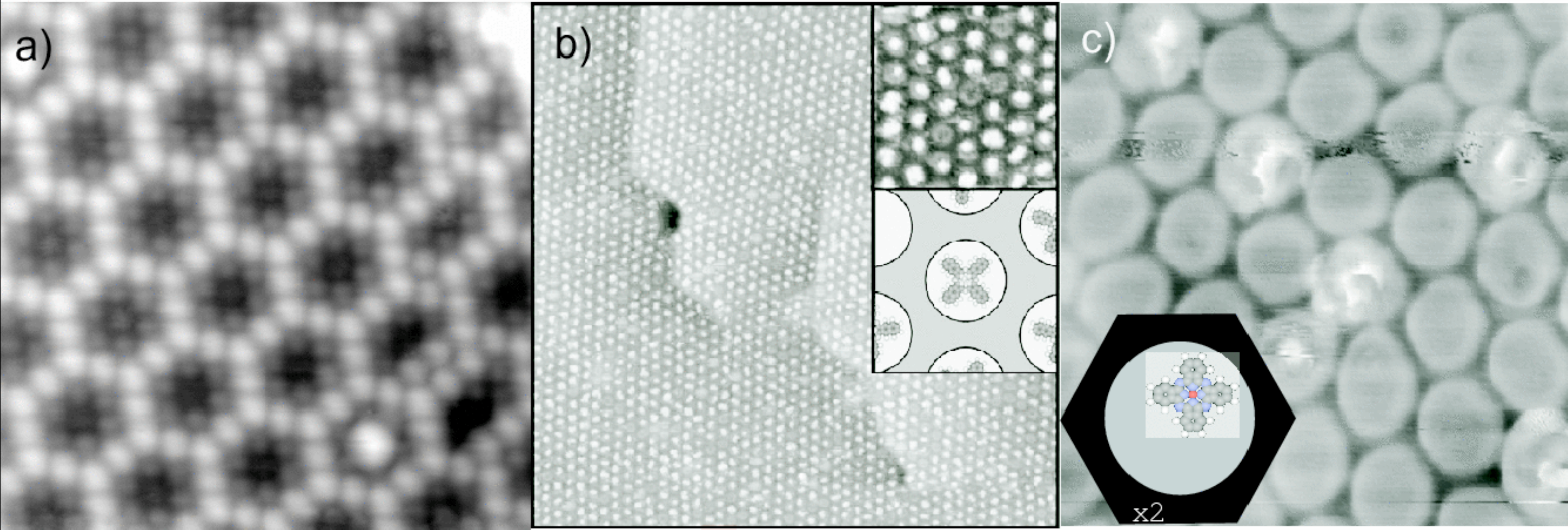}}
\caption{Room temperature scanning tunneling microscopy (STM) images of molecules trapped in h-BN/Rh(111) nanomesh.
a) C$_{60}$: Individual molecules are imaged throughout this region, following closely
the topography (15x15 nm). The positions in the hole centers are occupied by either zero or
one C$_{60}$ molecule; at two places, large protrusions may represent additional corralled molecules (From \cite{cor04}).
b) Naphthalocyanine (Nc) (C$_{48}$H$_{26}$N$_{8}$):  Site-selective adsorption (120x120 nm). The inset (19x19 nm) on the top right shows an enlargement. The inset on the right is a schematic representation of the molecule in the h-BN nanomesh. (From \cite{ber07}). 
c) Copper phthalocyanine (Cu-Pc) (C$_{32}$H$_{16}$CuN$_{8}$) molecules trapped at the rim of the holes (18 x 18 nm). At the given tunneling conditions, the mesh wires map dark, the holes map grey and Cu-Pc are imaged as bright objects. The inset shows a magnified model of the trapped Cu-Pc molecule. (From \cite{dil08}). }
\label{Ftemplate}
\end{figure}

Figure \ref{Ftemplate} documents the template function of h-BN/Rh(111) for three different molecules at room temperature.
For C$_{60}$ with a van der Waals diameter of about 1 nm it is found that h-BN is wetted, and 12 C$_{60}$ molecules sit in one Rh (12x12) unit cell.
It can also be seen that the ordering is not absolutely perfect.
There are super cells with one  C$_{60}$ missing and such where one extra  C$_{60}$ is "coralled" in the holes of the h-BN nanomesh.
If we take a molecule that has the size of the holes of 2 nm, here naphthalocyanine (Nc) (C$_{48}$H$_{26}$N$_{8}$), we observe that the molecules self assemble in the holes.
Apparently, the trapping potential is larger than the molecule-molecule interaction that would lead to the formation of Nc islands with touching molecules.
If we take a molecule with an intermediate size, i.e. copper phthalocyanine (Cu-Pc) (C$_{32}$H$_{16}$CuN$_{8}$) with a van der Waals diameter of 1.5 nm, it is seen that they also assemble in the nanomesh holes.
There is a significant additional feature compared to the Nc case, i.e. Cu-Pc does not sit in the center of the holes but likes to bind at the rims. 
This observation gives an important hint to the bond energy landscape within this superstructure.

\begin{figure}
\centerline {\includegraphics[width=0.8\textwidth]{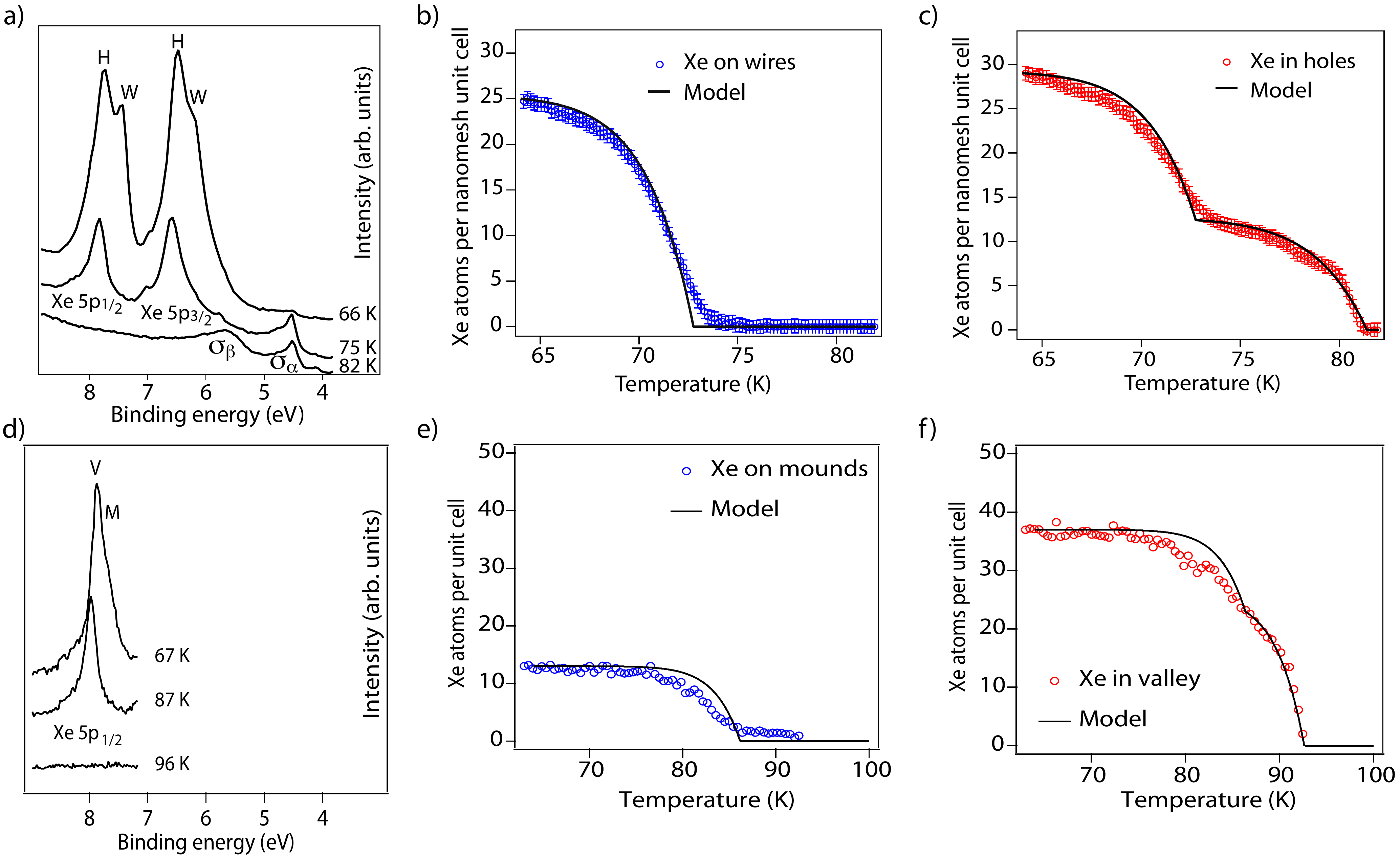}}
\caption{Temperature dependence of normal emission valence band photoemission spectra from Xe/h-BN/Rh(111) \cite{dil08} and Xe/g/Ru(0001) \cite{bru09}. a) Energy distribution curves extracted for three different temperatures (66, 75, and 82 K) for Xe/h-BN/Rh(111). b) Spectral weight of the Xe on the wires as a function of temperature.  c) Spectral weight of the Xe in the holes as a function of temperature.  d) Energy distribution curves extracted for three different temperatures (67, 87, and 96 K) for Xe/g/Ru(0001). e) Spectral weight of the Xe on the mounds as a function of temperature. f) Spectral weight of the Xe in the valleys as a function of temperature.  
The solid lines in b), c), e) and f) are fits obtained from a zero-order desorption model.}
\label{FXe}
\end{figure}
In Section \ref{lateraltextured} it was shown that corrugated $sp^2$ layer systems have relatively large lateral electric fields due to dipole rings.
These fields may polarize molecules and provide an additional bond energy.
Figure \ref{F3} shows that this bonding scales with $\alpha \cdot E^2_\parallel$, where $E_\parallel$ is the lateral electric field, and $\alpha$ the polarizability of the molecule.
$E_\parallel ^2$ is largest at the rims of the h-BN nanomesh holes or the g/Ru(0001) mounds.
This feature in the bond energy landscape is in line with the observation that molecules like Cu-Pc like to sit at the corrugation rims of the $sp^2$ layers. 
A quantitative measure for the adsorption energy can be obtained from thermal desorption spectroscopy (TDS) which was performed for xenon.

\begin{table}
\begin{tabular}{|l|r|r|r|r|r|r|}
\hline
&\multicolumn{3}{c|}{$h$-BN/Rh(111)}&\multicolumn{3}{c|}{g/Ru(0001)}\\
\hline
Phase&$C^W$&$C^H$&$R^H$&$C^M$&$C^V$&$R^V$\\
\hline
$E_d$(meV)&$181$&$184$&$208$&$222$&$222$&$234$\\
$N_1$&25&17&12&13&14&23\\
\hline
\end{tabular}
\caption{Experimentally determined Xe desorption energies  $E_d$ and Xe atoms per unit cell at full coverages $N_1$ for $h$-BN/Rh(111) \cite{dil08} and g/Ru(0001) \cite{bru09}. For all fits an attempt frequency $\nu$ of ${1.2\times10^{12}}$ Hz has been used.}
\label{T2} 
\end{table}
Figure \ref{FXe} shows a comparison of Xe/h-BN/Rh(111) and Xe/g/Ru(0001) TDS, where the surfaces are heated with a constant heating rate $\beta=dT/dt$, and where the remaining Xe on the surface was monitored with photoemission from adsorbed xenon (PAX).
Since the Xe core level energy is sensitive to the local electrostatic potential it also indicates where it is sitting in the $sp^2$ super cell. 
This allows a TDS experiment where the Xe bond energy on different sites in the super cell is inferred \cite{dil08,bru09}. 

The desorption data for Xe/h-BN/Rh(111) on the wires and in the holes are shown in Figure \ref{FXe} b) and c), respectively, and for Xe/g/Ru(0001) on the mounds and the valley in Figure \ref{FXe} e) and f).
The data of the weakly bound $sp^2$ layer regions (wires and mounds) are well described with a zero order desorption model where $dN$ molecules desorb on the temperature increase $dT$:
\begin{equation}
-dN=\frac{\nu}{\beta}\cdot \exp(-\frac{E_d}{k_BT})\cdot dT
\label{zeroorder} 
\end {equation}

where $\nu$ is the attempt frequency in the order of 10$^{12}$ Hz, $\beta$ the heating rate, $E_d$ the desorption energy and $k_BT$ the thermal energy.
The data are fitted to the integrals of Equation \ref{zeroorder} where the initial coverage $N_1$ is taken from the intensity of the photoemission peaks, the sizes of the Xe atoms and the super cells. 
From this it is e.g. found that 25 Xe atoms cover the wires in the h-BN/Rh(111) unit cell.
The desorption energies of 169, 181, 222 and 249 meV for Xe/Xe \cite{ker05},  Xe$^W$/h-BN/Rh(111), Xe$^M$/g/Ru(0001) and for Xe/graphite \cite{ulb06} indicate that they are similar, but have a trend to increase in going to more metallic substrates. 
The initial coverages and desorption energies are summarized in Table \ref{T2}.
For the strongly bound regions (hole and valley) this single desorption energy picture does not hold.
For Xe/h-BN/Rh(111) it turned out that 12 Xe atoms in the holes have to be described with a larger bond energy. 
Correspondingly different phases $C$ and $R$ have been introduced, where $C$ stands for coexistence and $R$ for ring or rim \cite{dil08}.
Twelve Xe atoms fit on the rim of the h-BN/Rh(111) nanomesh holes and thus a dipole ring induced extra bond energy of 13\% was inferred.
For g/Ru(0001) the situation is less clear cut, since a two phase fit fits the data less well than those of the h-BN/Rh(111) case. 
This may be due to the different topography of the two systems (see Figure \ref{F1Brugger}). The fact that the extra bond energy in Xe/g/Ru(0001) is 4\% only is rationalized with the lower local work function difference, i.e. the smaller Xe core level energy splitting in Xe/g/Ru(0001) (240 meV) than in Xe/h-BN/Rh(111) (310 meV).

%%%%%%%%%%%%
%%%%%%%%%%%%
\newpage
\appendix\section{Atomic and Electronic Structure in Real and Reciprocal Space}

\begin{figure} [!b]
\centerline{ \includegraphics[width=0.5\textwidth]{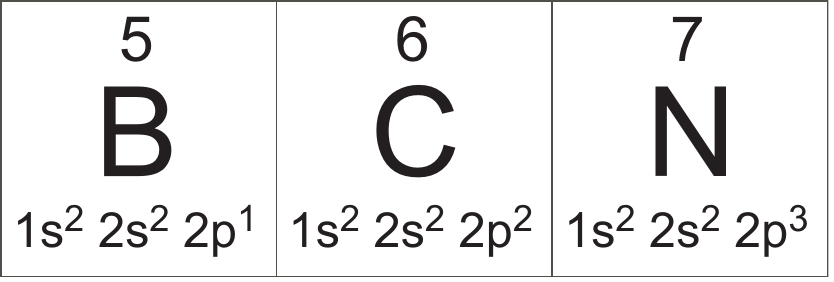} }
\caption{Boron, carbon and nitrogen are neighbors in the periodic table. Accordingly, h-BN and graphene are isoelectronic and are constituted by 12 electrons per unit cell.}
\label{FigBCN}
\end{figure}

Graphene and hexagonal boron nitride layers are isoelectronic.
Both, the (C,C) and the (B,N)  building blocks have 12 electrons and 
both have a honeycomb network structure with fairly strong bonds of 6.3 and 4.0 eV for C=C and B=N, respectively.
The electronic configurations of the constituent atoms are shown in Figure \ref{FigBCN}.

%These strong bonds, indicate high stability, which is an important ingredient for their usefulness as robust materials.
 
\subsection{$sp^2$ hybridization} 

\begin{figure}
\centerline{ \includegraphics[width=0.5\textwidth]{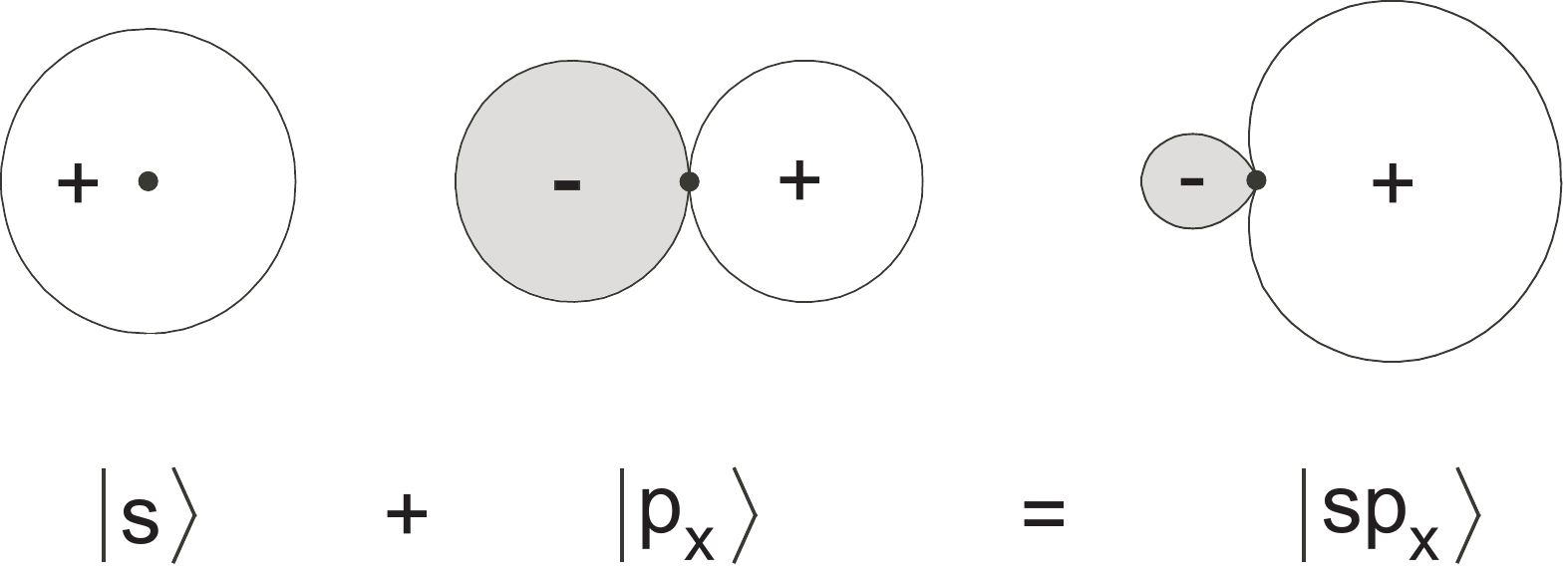} }
\caption{The coherent sum of a $\vert s\rangle$ and a $\vert p_x\rangle$ wavefunction leads to a directional $\vert sp^2\rangle$ hybrid. The phases of the wavefunctions are marked with + and -, where negative amplitudes are shaded.}
\label{sphybrid}
\end{figure}

The strong bonding within the $sp^2$ network foots on the hybridization of the 2s and the 2p valence orbitals.
The 2s and the 2p energies in atomic  boron, carbon and nitrogen are close in energy (in the order of 10 eV) and have similar spacial extension.
If the atoms are assembled into molecules, where the overlap between adjacent bonding orbitals is maximised, the linear combination of the s and p orbitals
provides higher overlap than that of two 2p orbitals. 
Figure \ref{sphybrid} shows a $s$ and a $p_x$ wavefunction and their  $sp_x$ hybrid, which is a coherent sum of the wavefunction amplitudes. 
The phase in the $p$ wavefunction depends on the direction and produces the lobe along the $x$ direction, which allows a large overlap with the wave functions of the neighboring atoms.

The hybrid orbitals are obtained as a linear combination of atomic orbitals. 
For the case of carbon $sp^3$, $sp^2$ and $sp^1$ hybrids may be formed, where CH$_4$ (methene), C$_2$H$_4$ (ethene) and C$_2$H$_2$ (acetylene)
are the simplest hydrocarbons representing these hybrids.
The tetrahedral symmetry of $sp^3$ and the planar symmetry of $sp^2$ is also reflected in the two allotropes diamond and graphite, where graphite is the thermodynamically stable phase. 
For boron nitride, accordingly cubic and hexagonal boron nitrides are found, where here the cubic form is thermodynamically more stable at room temperature and 1 bar pressure.

The $sp$ hybrid orbitals are obtained in combining $s$ and $p$ orbitals.
This corresponds to a new base for the assembly of atoms into molecules.
In the $sp^2$ hybrid orbitals the index 2 indicates that two $p$ orbitals are mixed with the $s$ orbital.
The $2s$ orbital is combined with $2p_x$ and $2p_y$ orbitals, into 3 two fold spin degenerate orbitals that form the $\sigma$ bonds:

\begin{equation}
\vert sp^2_1\rangle={\sqrt \frac{1}{3}} \vert s\rangle+{\sqrt \frac{2}{3}}\vert p_x\rangle
\end {equation}

\begin{equation}
\vert {sp^2_2\rangle={\sqrt \frac{1}{3}} \vert s\rangle- {\sqrt \frac{1}{6}} \vert p_x\rangle + {\sqrt \frac{1}{2}} \vert p_y\rangle}
\end {equation}

\begin{equation}
\vert sp^2_3\rangle={\sqrt \frac{1}{3}} \vert s\rangle- {\sqrt \frac{1}{6}} \vert p_x\rangle - {\sqrt \frac{1}{2}}\vert p_y\rangle
\end {equation}

These 3 $\sigma$ orbitals contain a mixture of $\frac{1}{3}$ of an $s$ and $\frac{2}{3}$ of a $p$ electron.
It can easily be seen that the three orbitals lie in a plane and point in directions separated by angles of 120$^\circ$.
$\vert sp^2_1\rangle$ points into the direction [${\sqrt{\frac{2}{3}}},0,0$], $\vert sp^2_2\rangle$ into the direction [-$\sqrt{\frac{1}{6}},{\sqrt{\frac{1}{2}}},0$], and from the scalar product between the two directions, we get the angle of 120$^\circ$.
In the case of the $sp^2$ hybridization the fourth orbital has pure $p$ character and forms $\pi$ bonds.
\begin{equation}
\vert sp^2_4\rangle= \vert p_z\rangle
\end {equation}
$\vert sp^2_4\rangle$ is perpendicular to the $sp^2$ $\sigma$ bonding plane.
For the case of graphene these $p_z$ orbitals on the honeycomb $sp^2$ network are responsible for the spectacular electronic properties of the conduction electrons in the $\pi$ bands because they are occupied with one electron.

\subsection{Electronic band structure}
\label{Elbandstructure}

\begin{figure}
\centerline{ \includegraphics[width=0.6\textwidth]{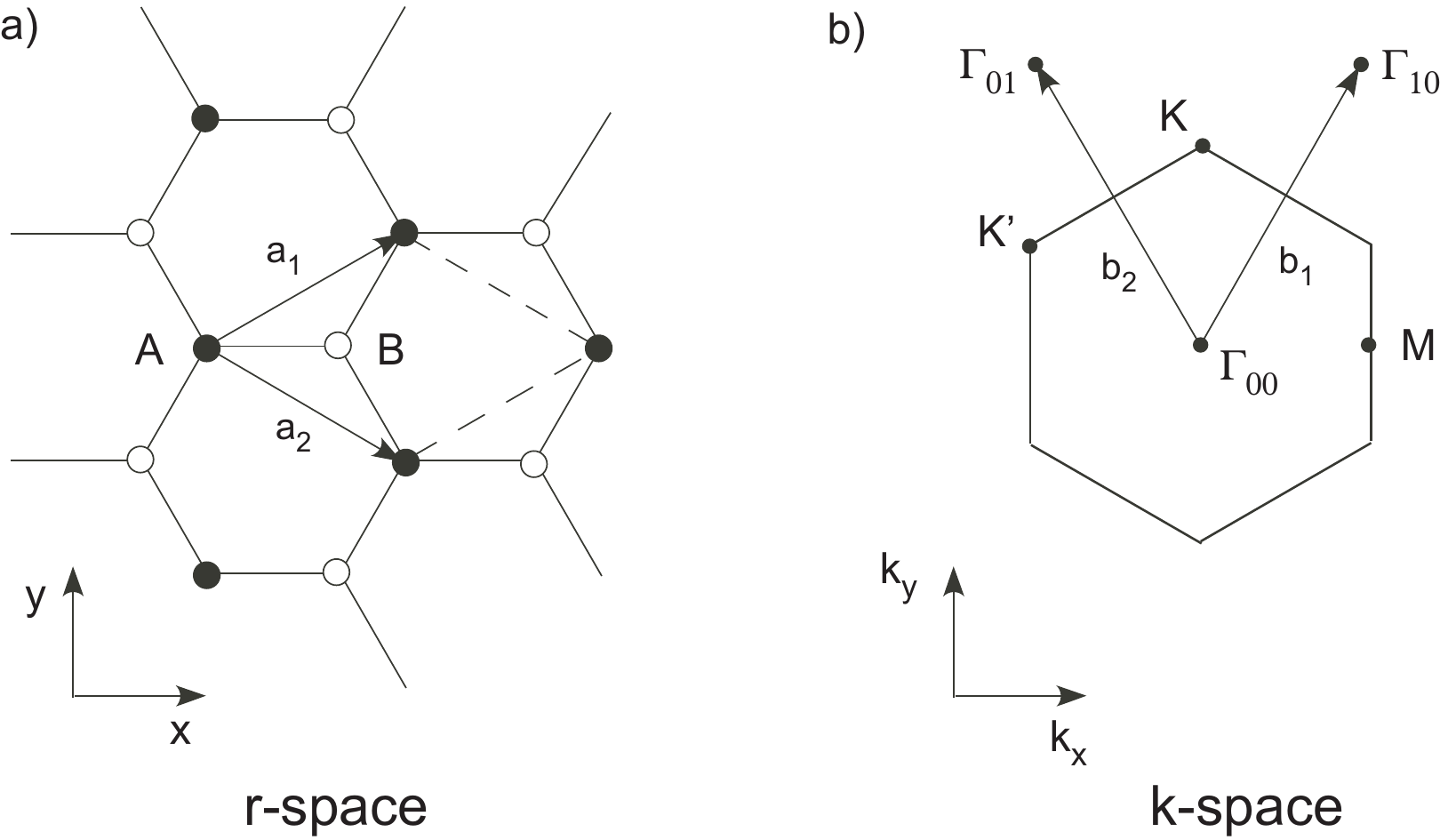} }
\caption{Honeycomb structure like that of a single layer h-BN or graphene. a) real space: the $sp^2$ hybridization causes the formation of two coupled sublattices $A$ and $B$ with lattice vectors ${\bf a}_1$ and  ${\bf a}_2$ and lattice constant $a=|{\bf{a}}_1|=|{\bf{a}}_2|$.  b) reciprocal space: Brillouin zone. The high symmetry points $\Gamma, M$ and $K$ are marked, where the reciprocal distance $\Gamma K$ is $\frac{2\pi}{\sqrt{3}a}$. The distinction of $K$ and $K'$ is possible if the system is three fold symmetric (trigonal), but not six-fold symmetric (hexagonal).}
\label{RKStructure}
\end{figure}

The $sp^2$ hybridization determines the {\bf {atomic structure}} of both, hexagonal boron nitride and graphene sheets. They  form a two dimensional honeycomb structure as shown in Figure \ref{RKStructure}.
The lattice can be described as superposition of two coupled sublattices $A$ and $B$ (see Figure \ref{RKStructure} a)).
In the case of graphene both sublattices are occupied by one carbon atom ($C_A,C_B$), while in the case of h-BN one sublattice is occupied by boron atoms and the other sublattice by nitrogen atoms (B,N). 
The interference of the electrons between these lattices causes the peculiar electronic structure of  $sp^2$ layer networks. 
The 12 electrons in the unit cell are filled  into 4 $1s$ core levels, and into the 16 $sp^2$ hybrids that form 3 $\sigma$ bonding, one $\pi$ bonding, one $\pi^*$ antibonding and 3$\sigma^*$ anti bonding bands. In the bonding bands the two adjacent $sp^2$ hybrids are in phase, while in the antibonding case they are not.
The atomic orbitals $sp^2_1$, $sp^2_2$ and $sp^2_3$ constitute the in plane $\sigma$ bands, while the $sp^2_4=p_z$ orbitals form the $\pi$ bands.
From this it can be seen that for flat layers the $\sigma$ and the $\pi$ electrons do not interfere, i.e. may be treated independently.
In Figure \ref{RKStructure} the real space lattice and the corresponding Brillouin zone in reciprocal space are shown. 
In $k$-space the reciprocal lattice constant is $\frac{2\pi}{a}$, where $a$ is the lattice constant of graphene or h-BN, of about 0.25 nm.
For the case of graphene, where the two sublattices are indistinguishable, this leads to a gap-less semiconductor with Dirac points at the $K$ points. 
In h-BN, the distinguishability of boron and nitrogen leads to an insulator, where the $\pi$ valence band is mainly constituted by the nitrogen sublattice, and the conduction band by the boron sublattice.

The different electronegativities of boron and nitrogen lead for free-standing h-BN to 0.56 $e^-$ transferred from B to N \cite{gra03}. 
This ionicity produces a Madelung energy $E_{Mad}$:
\begin{equation}
E_{Mad}=\alpha_{Mad}\cdot \frac{1}{4 \pi \epsilon_{0}} \cdot \frac{q^2}{a}
\label{Mad}
\end{equation}
with $\alpha_{Mad}$=1.336 for the honeycomb lattice \cite{roz96}, $q$ the displaced charge (in the above case 0.56 $e^-$) and $a$ the lattice constant.
It has to be noted that this Madelung energy applies for the free-standing case. 
If the ionic honeycomb layer sits on top of a metal, the energy in Equation \ref{Mad} reduces by a factor of 1/2.

\subsubsection{$\pi$ bands}

It is instructive to recall the basic statements within the framework of the  "tight binding" scheme of Wallace that he developed for the band theory of graphite  \cite {wal47}. 
Also the description of tight binding calculations of molecules and solids of Saito and Dresselhaus is recommended \cite{sai98} and the most recent review of Castro Neto et al. \cite{cas09}.
Tight binding means that we approximate the wave functions as superpositions of atomic $p_z$ wave functions on sublattice $A$ and $B$, respectively.
Furthermore, they are Bloch functions with the periodicity of the lattice.
The essential physics lies in the interference between the two sublattices. 
It is just another beautiful example for quantum physics with two interfering systems.
In order to solve the Schr\"odinger equation with this Ansatz we have to solve the secular equation:
\begin{equation}
\left |\begin{array}{cc}
H_{AA}-E & H_{AB}\\
H_{BA}& H_{BB}-E\\
\end{array} \right |
=0
\end{equation}
where the matrix elements $H_{AA}$ and $H_{BB}$ describe the energy on the sublattices $A$ and $B$ and most importantly $H_{AB}$ the hybridization energy due to interference or hopping of the $\pi$ electrons between the two lattices. 

The solutions for the energies $E_-$ (bonding) and $E_+$ (antibonding) are:

\begin{equation}
E_{\pm}=\frac{1}{2} \left (H_{AA}+H_{BB}\pm{\sqrt {(H_{AA}-H_{BB})^2+4|H_{AB}|^2}}\right)
\label{EqE}
\end{equation}

Here we show the simplest result that explains the essential physics.
It is the degeneracy of the bonding $\pi$ and the antibonding $\pi^*$ band at the $K$ point of the Brillouin zone, if the two sublattices are indistinguishable. 
This means that at $K$ the square root term in Equation \ref{EqE} has to vanish.
%If e.g. the overlap between the $p_z$ orbitals would be considered Equation \ref{} has tobe normalized with an overlap term.

The tight binding ansatz delivers values for $H_{AA}$, $H_{BB}$ and $H_{AB}$:

\begin{equation}
%H_{AA}= E_A-2\gamma_{AA} \left(2 \cos(\sqrt{3} \pi k_x a)\cos(\pi k_y a)+\cos(2\pi k_y a)\right)
% beachte k-kalibration .
H_{AA}= E_A
\end{equation}

\begin{equation}
%H_{BB}= E_B-2\gamma_{BB} \left(2 \cos(\sqrt{3} \pi k_x a)\cos(\pi k_y a)+\cos(2\pi k_y a)\right)
% beachte k-kalibration .
H_{BB}= E_B
\end{equation}

where $E_A$ and $E_B$ are the unperturbed energies of the atoms on sublattices $A$ an $B$.
The square of the interference term $H_{AB}$ is $k$-dependent and gets:

\begin{equation}
\left|H_{AB} \right|^2= \gamma_{AB}^2\left(1+4\cos(k_x a) \cos(k_y a/\sqrt{3})+ 4\cos^2( k_y a/\sqrt{3})\right)
\label{HAB}
\end{equation}

$k_x$ and $k_y$ are coordinates in $k$-space pointing along $x$ and $y$, respectively, and $a$ is the lattice constant (see Figure \ref{RKStructure}). $\gamma_{AB}$ describes the hybridization between the sublattices and is proportional to the electron hopping rate between two adjacent sites on sublattice $A$ and sublattice $B$. 
Basically it determines the $\pi$ band width, which turns out to be $3\gamma_{AB}$.
If hopping within the sublattices, e.g. between two adjacent A-sites, is allowed this leads to $k$-dependent corrections in $H_{AA}$ and $H_{BB}$ and to a symmetry breaking between the $\pi$ and the $\pi^*$ band \cite{wal47}. If the phase of the electron wave functions is considered, the hexagonal symmetry is broken and a trigonal symmetry, i.e. two distinct $K$ points, $K$ and $K'$ have to be considerd \cite {mcc56,cas09}.
For equivalent sublattices, i.e.  $H_{AA}$ = $H_{BB}$ the $\pi$ bandstructure is given by the hopping between the two sublattices i.e. by $H_{AB}$ and $H_{BA}$, respectively. 
From Equation \ref{HAB} it is seen that $H_{AB}$ vanishes at the $K$ point, i.e. for ${\bf{k}}=(k_x,k_y)=(0,\frac{2\pi}{{\sqrt{3}}a})$.
The fact that the electrons in the two sublattices do not interfere if they are at the $K$ point of the Brillouin zone, is a direct consequence of the symmetry of the crystal, and does not change if hopping between non nearest neighbors is included in the model.
It should also be mentioned that these results are expected for the electronic structure of any system that forms a honeycomb lattice.

\begin{figure}
\centerline{ \includegraphics[width=0.9\textwidth]{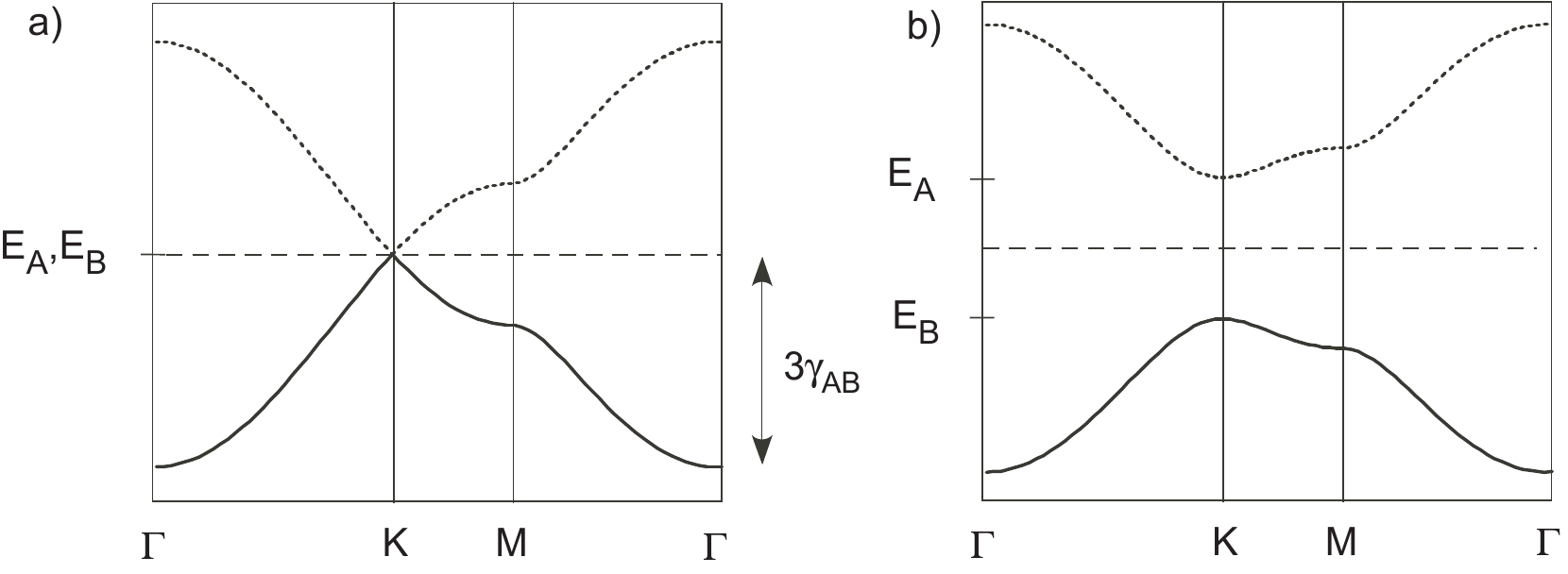} }
\caption{Tight binding $\pi$ and $\pi^*$ band structures of honeycomb lattices as shown in Figure \ref{RKStructure} along $\Gamma K M \Gamma$. The hopping rate $\gamma_{AB}/\hbar$ is kept constant for both cases. a) Case for two undistinguishable sublattices $A$ and $B$ with $E_A$=$E_B$. Note the emergence of a gapless semiconductor: If each sublattice contributes one electron, the Fermi surface is constituted by the Dirac point at $K$. b) Case for two distinguishable sublattices $A$ and $B$ with $E_A>E_B$. Note: At $K$ a gap with the magnitude $E_g=E_A-E_B$ opens, and with an even number of $\pi$ electrons the system is an insulator.}
\label{GKMG}
\end{figure}

In Figure \ref{GKMG} the generic tight binding band structure of a $sp^2$ hybridized lattice with two sublattices is shown for hopping $\gamma_{AB}$ between the sublattice A and B, only. 
Figure \ref{GKMG} a) presents the case of graphene, i.e. equivalent sublattices $A$ and $B$. 
The two $p_z$ electrons from the atoms $C_A$ and $C_B$ fill the $\pi$ band.
The highest energy corresponds to the Fermi energy and lies at the $K$ point. 
This means that the Fermi surface of this system consists of points at the $K$ points. 
The peculiarity that the resulting Fermi surface encloses no volume leads to the notion that graphene is a gapless semiconductor.
The band structure in the vicinity of  the $K$ point is most interesting.
The dispersion $E(k)$ of the electrons is linear, i.e. $\left(\frac{\partial E}{\partial k} \right)_K=v_F=const.$. 
The linear dispersion resembles to that of massless photons, or relativistic particles with $E\gg m_oc^2$, where $m_o$ is the rest mass.
It constitutes so called Dirac cones at the $K$ point of the Brillouin zone (see Figure \ref{GKMG}) a)).
From Equation \ref{HAB} we get with $\frac{\partial{E}}{\partial{k}}$ at the $K$ point  the Fermi velocity  $v_F=\frac{a\cdot\gamma_{AB} }{\hbar}$ , which is for a $\pi$ bandwidth of 6 eV or hopping rate of $\frac{1}{10 \rm {fs}}$ about $c/300$.
In brackets a seeming contradiction has to be clarified: 
The second derivative of $E(k)$ at the  $K$ point is zero.
With the relation for the second derivative of $ \frac{\partial^2 E}{\partial k^2}=\frac{\hbar^2}{m^*}$, the effective mass $m^*$ of the electrons and the holes is infinite. i.e. electrons and holes at the $K$ point may not be accelerated, as it is the case for photons.
However, since the Fermi velocity is not zero, electrons have not to be accelerated in order to be transported.
%Of course this has interesting consequences for electronic applications, where this is a benefit, since such charge moves much faster.  
 
In Figure \ref{GKMG} b) the result for a lattice with two inequivalent sublattices that corresponds to the case of h-BN are shown.
For the sake of simplicity, the same hybridization $\gamma_{AB}$ has been chosen.
The symmetry between the $A$ and the $B$ lattice is broken, if the energy of the unperturbed atoms $E_A$ and $E_B$ are not the same (see Equation \ref{EqE}), 
which is obviously the case for boron and nitrogen.
The band structure is similar, but at the $K$ point a gap with the magnitude of $|E_A-E_B|$ opens, and no Dirac physics is expected.

\subsubsection{$\sigma$ bands}

The $\sigma$ bands form the strong bonds between the atoms in the $x-y$ plane. For flat layers they are orthogonal to the $\pi$ bands and can be treated independently.
The case is more involved than that of the $\pi$ bands since here the 3 atomic orbitals ($s$, $p_x$ and $p_y$) on the two sublattices give rise to 6 bands. 
Also the overlap between the different atomic orbitals gets larger and the $s$-$p$ mixing is a function of the $k$ vector. 
With the tight binding ansatz similar to Equations \ref{EqE}-\ref{HAB} 3 $\sigma$ bands were derived \cite{sai98}.
Essentially the secular equation is now the determinant of a $6\times6$ matrix leading to 6 bands.
At $\Gamma$ the lowest lying band, $\sigma_0$, has $s$ character and the two remaining  $\sigma$ bands, $\sigma_1$ and $\sigma_2$ are degenerate and mainly $p_x$ and $p_y$ derived.

It is interesting to note that also for the $\sigma$ bands the bandstructure forms cones, if the base of the honeycomb lattice is homonuclear (graphene). They are reminiscent to the Dirac cone, where the $\pi$ and the $\pi^*$ band touch. 
At $K$ and a binding energy of about 13 eV the $\sigma_0$ and the $\sigma_1$ band touch.
For heteronuclear bases in the honeycomb (h-BN) also a gap opens, as it is observed for the $\pi$ and $\pi^*$ band.
Of course, the conical band touching is less important for graphene and h-BN, since both involved $\sigma$ bands remain fully occupied. 
The measurement of this gap, nevertheless, would open a way to distinguish the $\sigma$ and the $\pi$ bonding to the substrate.

\begin{figure}
\centerline{ \includegraphics[width=0.9\textwidth]{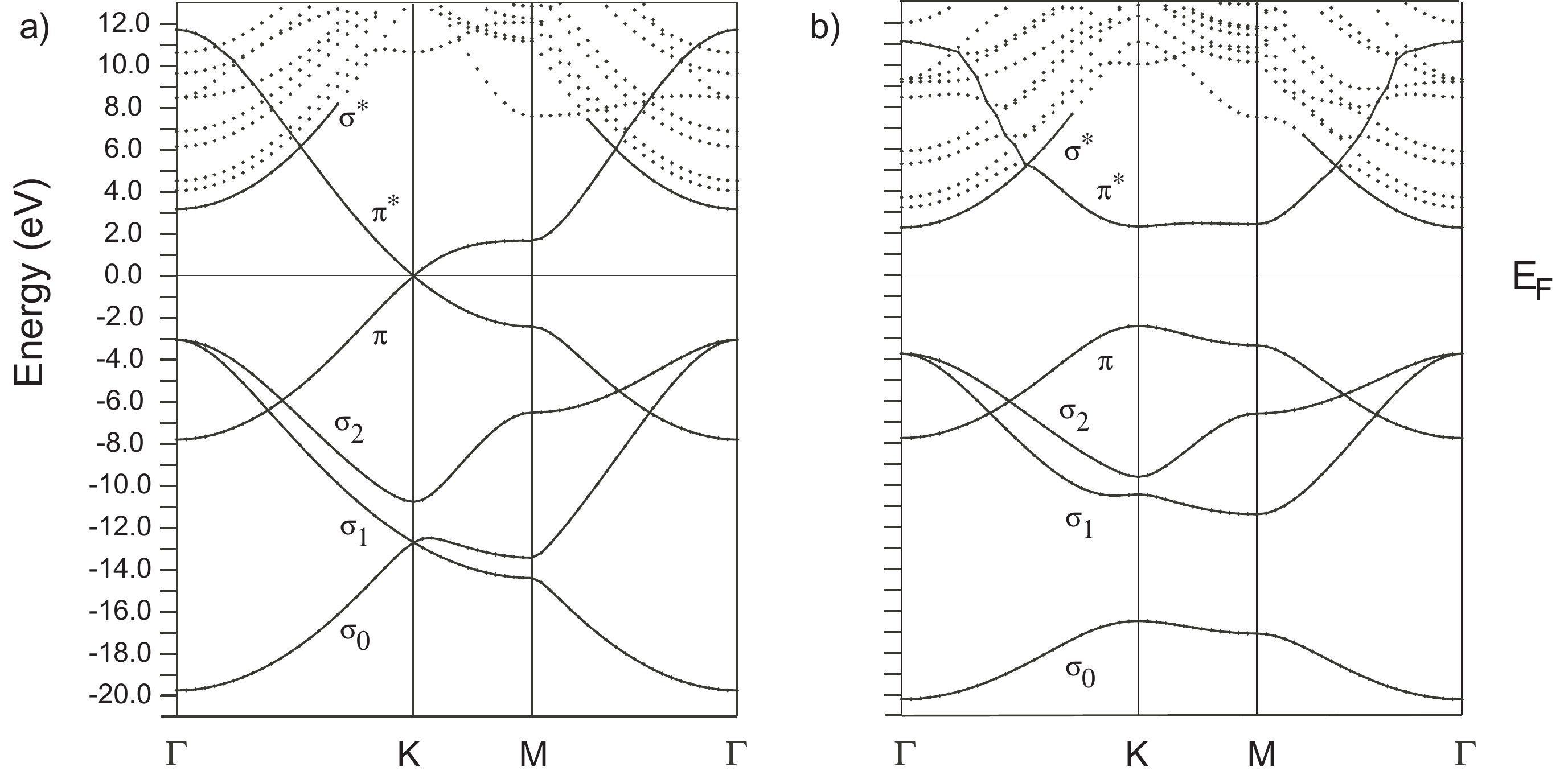} }
\caption{Density functional theory (DFT) bandstructure calculation along $\Gamma K M\Gamma$ for graphene and h-BN single layers. The eigenvalues for discrete $k$-points are shown as dots. The solid lines connect the eigenvalues in the Brillouin zone where the band assignment from the tight binding ansatz  with a 2s and 2p basis set holds: Three $\sigma$ bands $\sigma_0$, $\sigma_1$ and $\sigma_2$ and the  two $\pi$ bands $\pi$ and $\pi^*$ are shown. The three $\sigma^*$ antibonding bands are more difficult to resolve since they are also mixing in the DFT calculation with 3s and  3p contributions. a) Graphene. Note the Dirac cone at $K$ at the Fermi level $E_F$, where the $\pi$ and the $\pi^*$ bands touch.  For the $\sigma_0$ and the $\sigma_1$ band a similar conical band touching occurs at about -13 eV. b) h-BN. Note the band narrowing and the opening of a gap at the $K$ point for the $\pi$-$\pi^*$ and the $\sigma_0$-$\sigma_1$ band cones.
Calculations by courtesy of Peter Blaha.}
\label{Fsigma}
\end {figure}

Figure \ref{Fsigma} shows a state of the art  Density Functional Theory (DFT) bandstructure calculation for a single layer graphene and h-BN, respectively.
These calculations consider a much larger basis set than the 2s and the 3 2p orbitals as it is done in the tight binding picture. 
Accordingly more bands are found. 
The dots in Figure \ref{Fsigma} are the energy eigenvalues for given $k$ vectors. 
The three $\sigma$, the $\pi$ and the $\pi^*$ band reproduce the tight binding result \cite{sai98}. At higher energies also 3s and 3p orbitals contribute eigenvalues and the identification of the  antibonding bands becomes difficult.
\\

{\bf Acknowledgements}
\\

Most material presented in this Chapter was obtained with J\"urg Osterwalder and thanks to the empathy and work of our students Wilhelm Auw\"arter, Matthias Muntwiler, Martina Corso and Thomas Brugger. It is also a big pleasure to acknowledge Peter Blaha, who started in an early stage of our endeavor to contribute significantly with theory to the understanding of $sp^2$ single layers.

\newpage

\bibliographystyle{unsrt}

\end{document}